%% file: main.tex
\documentclass[12pt,a4paper]{article}

\usepackage{ifthen} % for conditional statements
\newboolean{pdflatex}
\setboolean{pdflatex}{true} % False for eps figures 

\newboolean{articletitles}
\setboolean{articletitles}{true} % False removes titles in references

\newboolean{uprightparticles}
\setboolean{uprightparticles}{false} %True for upright particle symbols

\newboolean{inbibliography}
\setboolean{inbibliography}{false} %True once you enter the bibliography

\def\paperauthors{Tim Gershon$^1$, Thomas Latham$^1$, Andy Morris$^2$, Wenbin Qian$^3$,\\Mark Whitehead$^4$, Ao Xu$^5$} 
\def\papertitle{Visualisation of \CP-violation effects in decay-time-dependent analyses of multibody $B$-meson decays} % Latex formatted title

\input{preamble}
\usepackage{relsize} % for mathlarger/mathsmaller
\usepackage[hang,flushmargin]{footmisc} 
\usepackage{comment}

\def\sintb{\ensuremath{\sin(2\beta)}\xspace}
\def\costb{\ensuremath{\cos(2\beta)}\xspace}
\def\sintbs{\ensuremath{\sin(2\beta_s)}\xspace}
\def\costbs{\ensuremath{\cos(2\beta_s)}\xspace}

\newcommand{\appropto}{\mathrel{\vcenter{
  \offinterlineskip\halign{\hfil$##$\cr
    \propto\cr\noalign{\kern2pt}\sim\cr\noalign{\kern-2pt}}}}}

\begin{document}

\renewcommand{\thefootnote}{\fnsymbol{footnote}}
\setcounter{footnote}{1}

\input{title}

\renewcommand{\thefootnote}{\arabic{footnote}}
\setcounter{footnote}{0}

\pagestyle{plain} % restore page numbers for the main text
\setcounter{page}{1}
\pagenumbering{arabic}

\input{paper}
\input{ack}

\addcontentsline{toc}{section}{References}
\setboolean{inbibliography}{true}
\bibliographystyle{LHCb}
\bibliography{references,standard,main,LHCb-PAPER,LHCb-CONF,LHCb-DP,LHCb-TDR}

\end{document}

%% file: preamble.tex
\usepackage{microtype}
\usepackage{lineno}  % for line numbering during review
\usepackage{xspace} % To avoid problems with missing or double spaces after
\usepackage{caption} %these three command get the figure and table captions automatically small

\usepackage{graphicx}  % to include figures (can also use other packages)
\usepackage{color}
\usepackage{colortbl}
\graphicspath{{./figs/}} % Make Latex search fig subdir for figures

\usepackage{amsmath} % Adds a large collection of math symbols
\usepackage{amssymb}
\usepackage{amsfonts}
\usepackage{upgreek} % Adds in support for greek letters in roman typeset

\newcommand*\patchAmsMathEnvironmentForLineno[1]{%
\expandafter\let\csname old#1\expandafter\endcsname\csname #1\endcsname
\expandafter\let\csname oldend#1\expandafter\endcsname\csname
end#1\endcsname
 \renewenvironment{#1}%
   {\linenomath\csname old#1\endcsname}%
   {\csname oldend#1\endcsname\endlinenomath}%
}
\newcommand*\patchBothAmsMathEnvironmentsForLineno[1]{%
  \patchAmsMathEnvironmentForLineno{#1}%
  \patchAmsMathEnvironmentForLineno{#1*}%
}
\AtBeginDocument{%
\patchBothAmsMathEnvironmentsForLineno{equation}%
\patchBothAmsMathEnvironmentsForLineno{align}%
\patchBothAmsMathEnvironmentsForLineno{flalign}%
\patchBothAmsMathEnvironmentsForLineno{alignat}%
\patchBothAmsMathEnvironmentsForLineno{gather}%
\patchBothAmsMathEnvironmentsForLineno{multline}%
\patchBothAmsMathEnvironmentsForLineno{eqnarray}%
}

\usepackage{hyperref}    % Hyperlinks in references
\usepackage[all]{hypcap} % Internal hyperlinks to floats.

\input{lhcb-symbols-def} % Add in the predefined LHCb symbols

\usepackage{cite} % Allows for ranges in citations
\usepackage{mciteplus}

%% file: lhcb-symbols-def.tex
\usepackage{xspace} 
\usepackage{upgreek}

\def\MagUp {\mbox{\em Mag\kern -0.05em Up}\xspace}

\ifthenelse{\boolean{uprightparticles}}%
{

 \def\Ppi         {\ensuremath{\uppi}\xspace}

 \def\Ppsi        {\ensuremath{\uppsi}\xspace}

 \def\PDelta      {\ensuremath{\Delta}\xspace}                 
 \def\PXi         {\ensuremath{\Xi}\xspace}                 
 \def\PLambda     {\ensuremath{\Lambda}\xspace}                 
 \def\PSigma      {\ensuremath{\Sigma}\xspace}                 
 \def\POmega      {\ensuremath{\Omega}\xspace}                 
 \def\PUpsilon    {\ensuremath{\Upsilon}\xspace}
 \let\oldPi\Pi
 \def\PPi         {\ensuremath{\oldPi}\xspace}

 \def\PB      {\ensuremath{\mathrm{B}}\xspace}                 
                  
 \def\PD      {\ensuremath{\mathrm{D}}\xspace}

 \def\PJ      {\ensuremath{\mathrm{J}}\xspace}                 
 \def\PK      {\ensuremath{\mathrm{K}}\xspace}

 \def\Pe      {\ensuremath{\mathrm{e}}\xspace}

 \def\Pi      {\ensuremath{\mathrm{i}}\xspace}

 \def\Ps      {\ensuremath{\mathrm{s}}\xspace}

 \def\thebaroffset{0.0em}
}
{

 \def\Ppi         {\ensuremath{\pi}\xspace}

 \def\Ppsi        {\ensuremath{\psi}\xspace}                 
                  
 \mathchardef\PDelta="7101
 \mathchardef\PXi="7104
 \mathchardef\PLambda="7103
 \mathchardef\PSigma="7106
 \mathchardef\POmega="710A
 \mathchardef\PUpsilon="7107
 \mathchardef\PPi="7105
                  
 \def\PB      {\ensuremath{B}\xspace}                 
                  
 \def\PD      {\ensuremath{D}\xspace}

 \def\PJ      {\ensuremath{J}\xspace}                 
 \def\PK      {\ensuremath{K}\xspace}

 \def\Pe      {\ensuremath{e}\xspace}

 \def\Pi      {\ensuremath{i}\xspace}

 \def\Ps      {\ensuremath{s}\xspace}

 \def\thebaroffset{0.18em}
}
\newcommand{\offsetoverline}[2][\thebaroffset]{\kern #1\overline{\kern -#1 #2}}%

\makeatletter
\ifcase \@ptsize \relax% 10pt
  \newcommand{\miniscule}{\@setfontsize\miniscule{4}{5}}% \tiny: 5/6
\or% 11pt
  \newcommand{\miniscule}{\@setfontsize\miniscule{5}{6}}% \tiny: 6/7
\or% 12pt
  \newcommand{\miniscule}{\@setfontsize\miniscule{5}{6}}% \tiny: 6/7
\fi
\makeatother

\DeclareRobustCommand{\optbar}[1]{\shortstack{{\miniscule (\rule[.5ex]{1.25em}{.18mm})}
  \\ [-.7ex] $#1$}}

\def\epem       {{\ensuremath{\Pe^+\Pe^-}}\xspace}
\def\squark    {{\ensuremath{\Ps}}\xspace}

\def\pion   {{\ensuremath{\Ppi}}\xspace}
\def\piz    {{\ensuremath{\pion^0}}\xspace}
\def\pip    {{\ensuremath{\pion^+}}\xspace}
\def\pim    {{\ensuremath{\pion^-}}\xspace}
\def\pipm   {{\ensuremath{\pion^\pm}}\xspace}

\def\kaon    {{\ensuremath{\PK}}\xspace}
\def\Kbar    {{\ensuremath{\offsetoverline{\PK}}}\xspace}

\def\KorKbar {\kern \thebaroffset\optbar{\kern -\thebaroffset \PK}{}\xspace}

\def\Kp      {{\ensuremath{\kaon^+}}\xspace}
\def\Km      {{\ensuremath{\kaon^-}}\xspace}

\def\KS      {{\ensuremath{\kaon^0_{\mathrm{S}}}}\xspace}

\def\KL      {{\ensuremath{\kaon^0_{\mathrm{L}}}}\xspace}

\def\Kstar   {{\ensuremath{\kaon^*}}\xspace}
\def\Kstarb  {{\ensuremath{\Kbar{}^*}}\xspace}

\def\Dbar    {{\ensuremath{\offsetoverline{\PD}}}\xspace}
\def\D       {{\ensuremath{\PD}}\xspace}

\def\DorDbar {\kern \thebaroffset\optbar{\kern -\thebaroffset \PD}\xspace}
\def\Dz      {{\ensuremath{\D^0}}\xspace}
\def\Dzb     {{\ensuremath{\Dbar{}^0}}\xspace}
\def\Dp      {{\ensuremath{\D^+}}\xspace}
\def\Dm      {{\ensuremath{\D^-}}\xspace}

\def\DpDm    {\ensuremath{\Dp {\kern -0.16em \Dm}}\xspace}

\def\B       {{\ensuremath{\PB}}\xspace}
\def\Bbar    {{\ensuremath{\offsetoverline{\PB}}}\xspace}

\def\BorBbar {\kern \thebaroffset\optbar{\kern -\thebaroffset \PB}\xspace}
\def\Bz      {{\ensuremath{\B^0}}\xspace}
\def\Bzb     {{\ensuremath{\Bbar{}^0}}\xspace}
\def\Bd      {{\ensuremath{\B^0}}\xspace}
\def\Bdb     {{\ensuremath{\Bbar{}^0}}\xspace}
\def\BdorBdbar {\kern \thebaroffset\optbar{\kern -\thebaroffset \Bd}\xspace}

\def\Bs      {{\ensuremath{\B^0_\squark}}\xspace}
\def\Bsb     {{\ensuremath{\Bbar{}^0_\squark}}\xspace}
\def\BsorBsbar {\kern \thebaroffset\optbar{\kern -\thebaroffset \Bs}\xspace}

\def\jpsi     {{\ensuremath{{\PJ\mskip -3mu/\mskip -2mu\Ppsi}}}\xspace}

\def\Y#1S{\ensuremath{\PUpsilon{(#1S)}}\xspace}

\def\LorLbar     {\kern \thebaroffset\optbar{\kern -\thebaroffset \PLambda}\xspace}

\def\to                 {\ensuremath{\rightarrow}\xspace}

\def\CP                {{\ensuremath{C\!P}}\xspace}

\def\AT#1     {\ensuremath{A_{\mathrm{T}}^{#1}}\xspace}           % 2

\def\C#1      {\ensuremath{\mathcal{C}_{#1}}\xspace}                       % 9
\def\Cp#1     {\ensuremath{\mathcal{C}_{#1}^{'}}\xspace}                    % 7
\def\Ceff#1   {\ensuremath{\mathcal{C}_{#1}^{\mathrm{(eff)}}}\xspace}        % 9  
\def\Cpeff#1  {\ensuremath{\mathcal{C}_{#1}^{'\mathrm{(eff)}}}\xspace}       % 7
\def\Ope#1    {\ensuremath{\mathcal{O}_{#1}}\xspace}                       % 2
\def\Opep#1   {\ensuremath{\mathcal{O}_{#1}^{'}}\xspace}                    % 7

\newcommand{\aunit}[1]{\ensuremath{\text{\,#1}}}       
\newcommand{\tev}{\aunit{Te\kern -0.1em V}\xspace}
\newcommand{\gev}{\aunit{Ge\kern -0.1em V}\xspace}
\newcommand{\mev}{\aunit{Me\kern -0.1em V}\xspace}
\newcommand{\kev}{\aunit{ke\kern -0.1em V}\xspace}
\newcommand{\ev}{\aunit{e\kern -0.1em V}\xspace}
 
\newcommand{\mevc}{\ensuremath{\aunit{Me\kern -0.1em V\!/}c}\xspace}
\newcommand{\gevc}{\ensuremath{\aunit{Ge\kern -0.1em V\!/}c}\xspace}
\newcommand{\mevcc}{\ensuremath{\aunit{Me\kern -0.1em V\!/}c^2}\xspace}
\newcommand{\gevcc}{\ensuremath{\aunit{Ge\kern -0.1em V\!/}c^2}\xspace}
\def\ps   {\ensuremath{\aunit{ps}}\xspace}

\def\gsim{{~\raise.15em\hbox{$>$}\kern-.85em
          \lower.35em\hbox{$\sim$}~}\xspace}
\def\lsim{{~\raise.15em\hbox{$<$}\kern-.85em
          \lower.35em\hbox{$\sim$}~}\xspace}

\def\sPlot{\mbox{\em sPlot}\xspace}

\def\degrees{\ensuremath{^{\circ}}\xspace}

\def\tell1  {TELL1\xspace}
\def\ukl1   {UKL1\xspace}

\newcommand{\eg}{\mbox{\itshape e.g.}\xspace}
\newcommand{\ie}{\mbox{\itshape i.e.}\xspace}

\newcommand{\lhcborcid}[1]{\href{https://orcid.org/#1}{\hspace*{0.1em}\raisebox{-0.45ex}{\includegraphics[width=1em]{figs/orcidIcon.pdf}}}}

%% file: title.tex
\begin{titlepage}
\pagenumbering{roman}

{\bf\boldmath\huge
\begin{center}
\papertitle
\end{center}}

\vspace*{1.5cm}

\begin{center}
\paperauthors
\bigskip\\
{\normalfont\itshape\footnotesize
$ ^1$University of Warwick, Coventry, United Kingdom\\
$ ^2$Aix Marseille Univ, CNRS/IN2P3, CPPM, Marseille, France\\
$ ^3$University of Chinese Academy of Sciences, Beijing, China\\
$ ^4$School of Physics and Astronomy, University of Glasgow, Glasgow, United Kingdom\\
$ ^5$INFN Sezione di Pisa, Pisa, Italy
}
\end{center}

\vspace{\fill}

\begin{abstract}
  \noindent
  Decay-time-dependent \CP-violation effects in transitions of neutral $B$ mesons to \CP-eigenstates can be visualised by oscillations in the asymmetry, as a function of decay time, between decay yields from mesons tagged as initially having $\Bbar$ or $\B$ flavour.
  Such images, for example for $B^0 \to \jpsi \KS$ decays where the magnitude of the oscillation is proportional to $\sin(2\beta)$ with $\beta$ being an angle of the Cabibbo--Kobayashi--Maskawa Unitarity Triangle, provide a straightforward illustration of the underlying physics.
  Until now there has been no comparable method to provide visualisation for the case of decays to multibody final states that are not \CP-eigenstates, where interference between \CP-even and -odd amplitudes provides additional physics sensitivity.  
  A method is proposed to weight the data so that the terms of interest can be projected out and used to obtain asymmetries that visualise the relevant effects.
  Application of the weighting to \Bs\ decays, where effects due to non-zero width difference are not negligible, provides a novel method to observe \CP\ violation in interference between mixing and decay without tagging the production flavour.
\end{abstract}

\vspace{\fill}

\end{titlepage}

\newpage
\setcounter{page}{2}
\mbox{~}

\clearpage

%% file: paper.tex
\section{Introduction}
\label{sec:intro}

\begin{figure}[tb]
    \centering
    \includegraphics[width=0.40\textwidth]{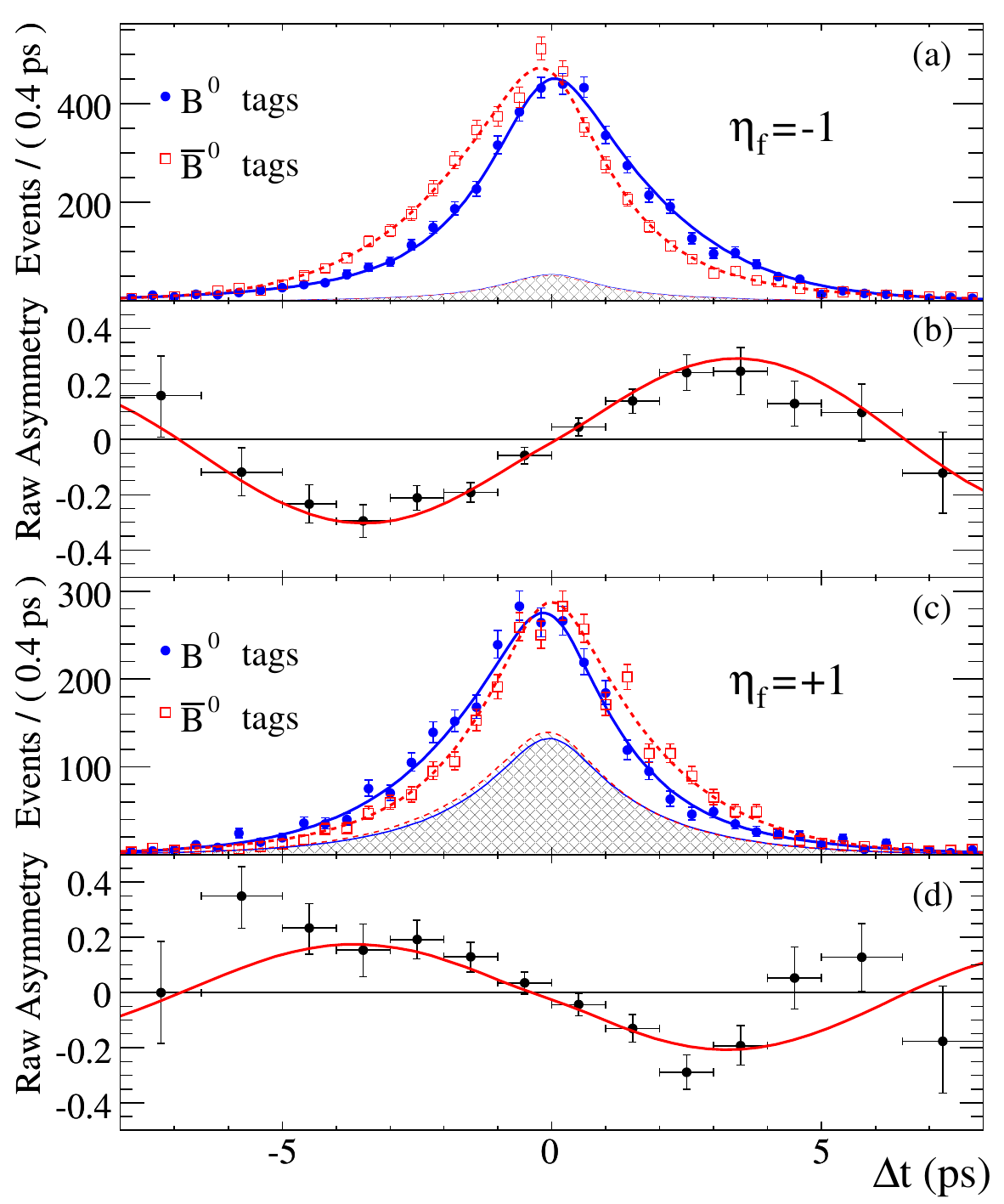}
    \hspace{3mm}
    \includegraphics[width=0.26\textwidth]{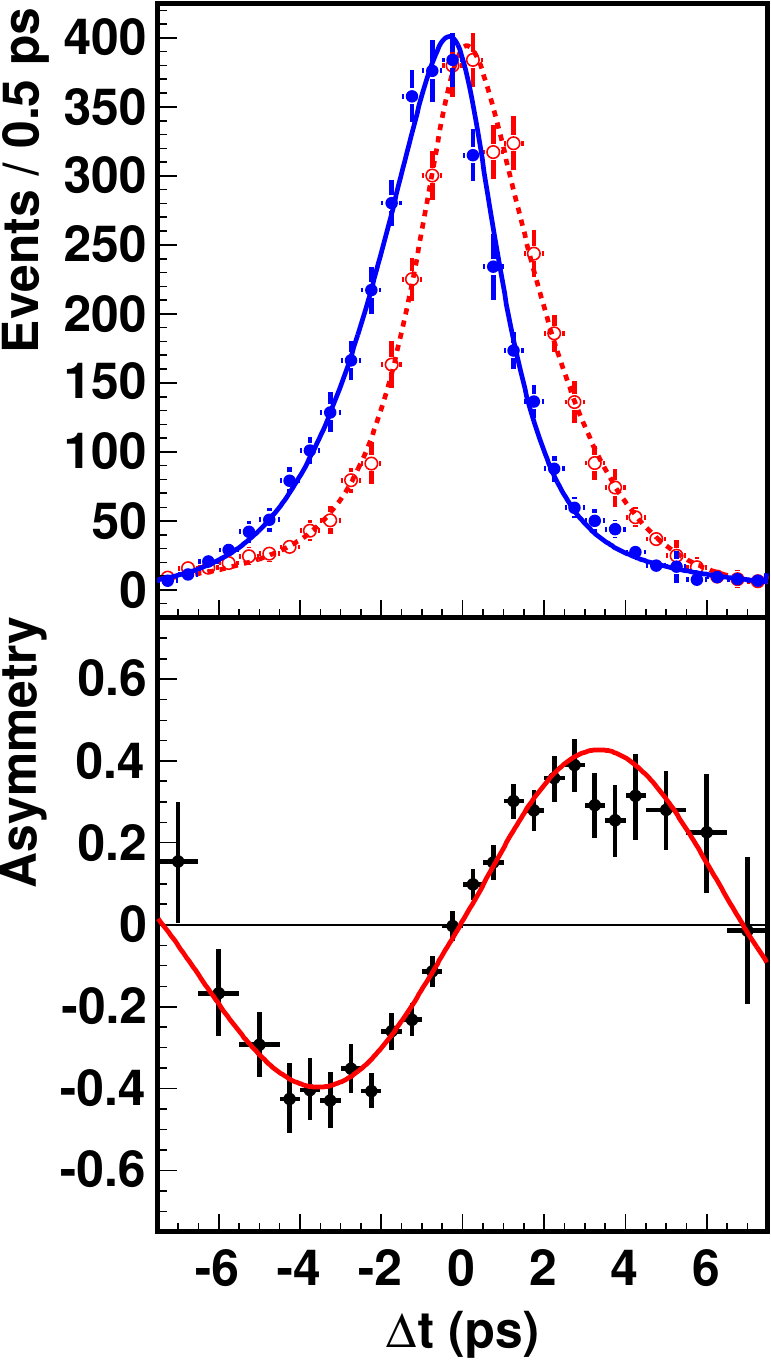}
    \includegraphics[width=0.26\textwidth]{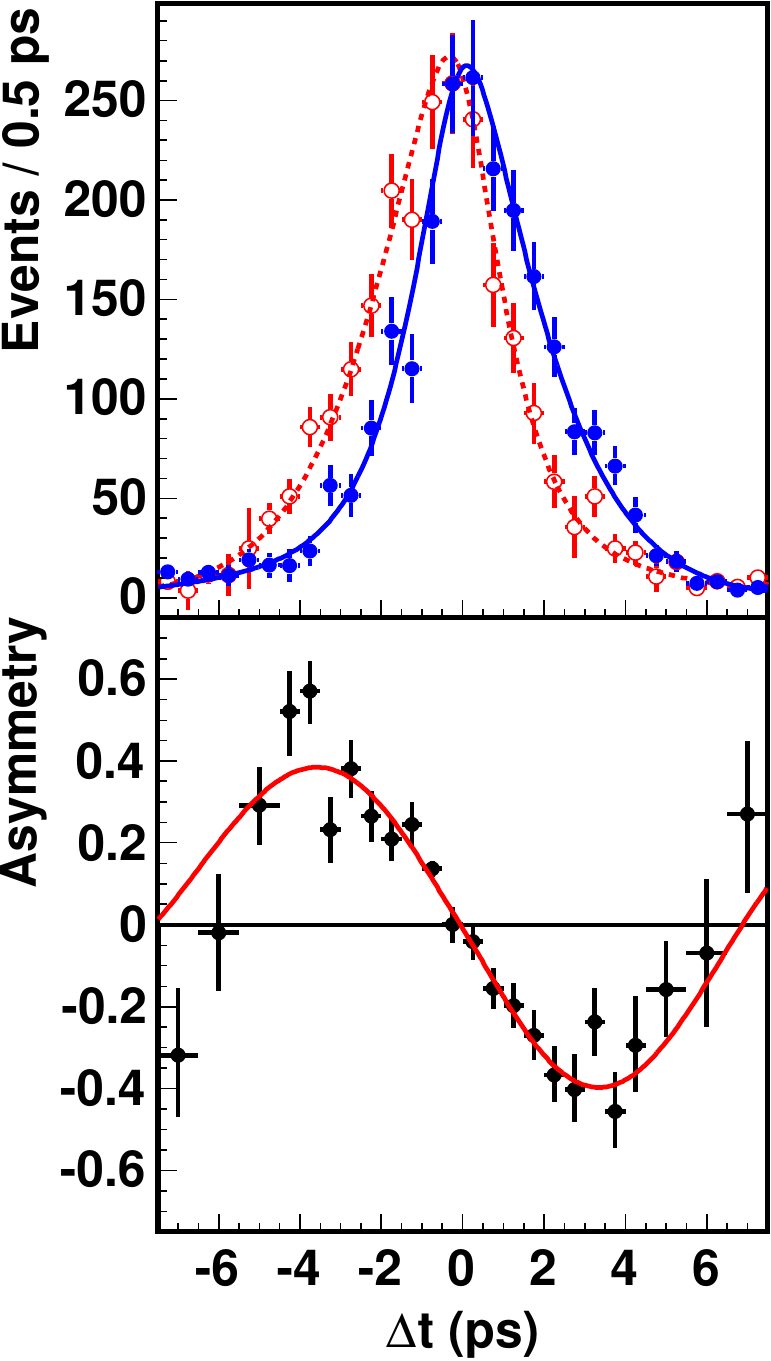}
    \caption{Decay-time distributions and asymmetries for $B^0 \to \jpsi \KS$ and similar decays from the (left)~BaBar~\cite{BaBar:2009byl} and (right)~Belle~\cite{Belle:2012paq} experiments.  In both cases the data are separated into (top for BaBar; left for Belle) \CP-odd final states such as $\jpsi \KS$ and (bottom for BaBar; right for Belle) \CP-even final states such as $\jpsi \KL$.
    Both experiments exploit $B$ production via the $\Upsilon(4S) \to \B\Bbar$ process, and hence use the time variable $\Delta t$, which is the difference in decay times of the signal and tagging \B meson in the $\B\Bbar$ pair.
    The data are separated into cases where the tagging \B meson is identified as having \Bz (blue for BaBar; red for Belle) or \Bzb (red for BaBar; blue for Belle) flavour; due to entanglement of the pair the signal \B meson is known to have the opposite flavour at the time of the tagging \B meson decay.  
    The BaBar data include background contributions, indicated by hatched regions, while the Belle data are background-subtracted.
    In both cases the asymmetry is diluted by imperfect flavour-tagging capability (the dilution is smaller for Belle since the figure includes only data with good quality tagging information).  
    }
    \label{fig:sin2beta-Bfactories}
\end{figure}

The topic of \CP\ violation is the subject of numerous investigations in contemporary particle physics.  
This includes many studies of decay-time-dependent \CP\ violation in transitions of neutral $B$ mesons to both \CP-eigenstates and non-\CP-eigenstates. 
A salient example is that of $B^0 \to \jpsi \KS$ decays, where the asymmetry in rates for neutral $B$ mesons, tagged as either $\Bzb$ or $\Bz$ at time $t=0$ and decaying at time $t$, can be written as 
\begin{equation}
    A_{\CP}(t) \equiv \frac{\Gamma[\Bzb \to \jpsi \KS(t)]-\Gamma[\Bz \to \jpsi \KS(t)]}{\Gamma[\Bzb \to \jpsi \KS(t)]+\Gamma[\Bz \to \jpsi \KS(t)]} = S \sin(\Delta m_d t) - C \cos(\Delta m_d t) \, ,
\end{equation}
where $\Delta m_d$ is the difference in masses between the heavier and lighter mass eigenstates of the $\Bz$--$\Bzb$ system.
In the limit that the $B^0 \to \jpsi \KS$ transition is dominated by the so-called tree amplitude, involving the Cabibbo--Kobayashi--Maskawa (CKM) quark-mixing matrix~\cite{Cabibbo:1963yz,Kobayashi:1973fv} elements $V^{}_{cb}V^{*}_{cs}$, it can be shown that $S = \sintb$ and $C=0$ to an excellent approximation~\cite{Carter:1980tk,Bigi:1981qs}, where $\beta$ is an angle of the CKM Unitarity Triangle.
Plots of this asymmetry in bins of decay-time were crucial for the BaBar and Belle experiments to demonstrate the existence of \CP\ violation in the \B-meson sector, with the final results from those experiments shown in Fig.~\ref{fig:sin2beta-Bfactories}~\cite{BaBar:2009byl,Belle:2012paq}.
%From fitting these data, BaBar obtain \mbox{$\sin(2\beta) = 0.687 \pm 0.028 \stat \pm 0.012 \syst$~\cite{BaBar:2009byl}} while Belle obtain \mbox{$\sin(2\beta) = 0.667 \pm 0.023 \stat \pm 0.012 \syst$~\cite{Belle:2012paq}}.

There are also numerous examples of \B meson decays to multibody final states that are not \CP-eigenstates, which are of interest to probe \CP violation effects.  
In such cases, interference between \CP-even and -odd amplitudes provides additional physics sensitivity.  
For example, the decay-time-dependent analysis of $\Bz \to \jpsi \Kstar(892)$, with $\Kstar(892) \to \KS\piz$,\footnote{
    The notation $\Kstar(892)$ is used to denote an admixture of the $\Kstar(892)^0$ and $\Kstarb(892)^0$ states, which is obtained since the final state, $\KS\piz$, does not determine the strangeness.
    Similarly, the notation $D^{(*)}$ will be used to denote a neutral $D^{(*)}$ meson that can be any admixture of $D^{(*)0}$ and $\overline{D}{}^{(*)0}$ states.
} decays can be used to determine \costb\ as well as \sintb, hence resolving a trigonometric ambiguity in the possible solutions for $2\beta$.
Such analyses have been performed by BaBar and Belle~\cite{BaBar:2004xhu,Belle:2005qtf}, although the results have large uncertainties.  
Decay-time-dependent amplitude analyses with sensitivity to both \costb\ and \sintb\ have also been carried out with the modes
$\Bd \to D^{(*)}h^0$ with $h^0 = \piz, \eta$ or $\omega$, $D^* \to D\piz$ and $D \to \KS\pip\pim$~\cite{BaBar:2018agf},
$\Bz \to \jpsi \pip\pim$~\cite{LHCb-PAPER-2014-058},
$\Bz \to \Kp\Km\KS$~\cite{BaBar:2012iuj,Belle:2010wis},
and $\Bz \to \KS\pip\pim$~\cite{BaBar:2009jov,Belle:2008til},
although in some of these cases the interpretation of the results in terms of CKM angles involves significant theoretical uncertainties.
In these publications, however, any visualisation of the data in terms of decay-time-dependent asymmetries is done by either integrating over the full phase-space of the decay or by selecting regions dominated by particular resonances.\footnote{
    An exception to this appears in Ref.~\cite{BaBar:2004xhu}, where Fig.~12 shows the asymmetry of data weighted by the moment of ${\cal C}$, where ${\cal C}$ is a function of angular variables and parameters in the decay $\Bz \to \jpsi \Kstar(892)$ with $\Kstar(892) \to \KS\piz$.
    This weighting allows the dependence of the data on \costb\ to be visualised, and  can be considered a specific implementation of the more general weighting approach proposed in this paper.
}
This tends to dilute, and can even completely remove, the dependence of the asymmetry on the physics parameters of interest.  
There are also additional methods, yet to be implemented in analysis of experimental data, that provide sensitivity to both \costb\ and \sintb, such as one using $\Bz \to D_{\CP}\pip\pim$ decays, where $D_{\CP}$ indicates that the neutral $D$ meson is reconstructed in a \CP\ eigenstate~\cite{Charles:1998vf,Latham:2008zs}.
There are further multibody $\Bz$ meson decay modes that provide sensitivity to other \CP-violation parameters, including $\Bz \to \pip\pim\piz$~\cite{Snyder:1993mx,BaBar:2013uwm,Belle:2007jkw,Belle:2007krm}.
Moreover, there are many \Bs\ decays of interest for similar reasons, including $\Bs \to \jpsi\Kp\Km$~\cite{LHCb-PAPER-2023-016,LHCb-PAPER-2017-008,ATLAS:2020lbz,CMS:2020efq}, $\jpsi \pip\pim$~\cite{LHCb-PAPER-2019-003}, $\phi\phi$~\cite{LHCb-PAPER-2023-001} and $\Kstar(892)^0\Kstarb(892)^0$~\cite{LHCb-PAPER-2017-048}.\footnote{
    Decays to two vector resonances, such as $\Bs \to \phi\phi$, are considered as having multibody final states for the purposes of this paper.
    As will become clear, the key point is that the final state involves a mixture of \CP-even and \CP-odd amplitudes.
}
As such, it is of interest to develop methods that enable the dependence of the data on the underlying physical parameters to be clearly visualised, with minimal dilution of the sensitivity.  

In the remainder of this paper, a method to achieve this goal is described.  
In Sec.~\ref{sec:methodology} the underlying methodology is set out, and an illustration of the method is given in Sec.~\ref{sec:illustration}.
The $\Bz \to D_{\CP}\pip\pim$ process is used as an example, but the method is general to any self-conjugate final state that is a mixture of \CP-even and \CP-odd contributions.  
This can be any multibody final state that contains a particle-antiparticle pair (\eg\ $\pip\pim$, $\Kp\Km$, $\Dz\Dzb$ and so on), since final states in which all particles are spin-0 \CP\ eigenstates are themselves \CP\ eigenstates~\cite{Gershon:2004tk}.
Approaches to handle experimental effects are discussed in Sec.~\ref{sec:experiment}. 
Uses of the method in the case of non-zero width difference, as is the case for \Bs-meson decays, are described in Sec.~\ref{sec:DeltaGamma}.
The impact of \CP\ violation in decay is considered in Sec.~\ref{sec:CPVinDecay}, and a summary concludes the paper in Sec.~\ref{sec:summary}.

\section{Methodology}
\label{sec:methodology}

The $\Bz \to D_{\CP}\pip\pim$ process~\cite{Charles:1998vf,Latham:2008zs} is used to illustrate the method.
If a neutral $B$ meson is tagged as \Bd at time $t=0$, and evolves at a later time $t$ into an admixture of $\Bz$ and $\Bzb$ that decays to a position $(m_+^2,m_-^2) \equiv \left( m^2(D\pip), m^2(D\pim) \right)$ in the $\D_{\CP} \pip \pim$ Dalitz plot, the corresponding amplitude is given by:
\begin{equation}
  A(m_+^2,m_-^2,t) = \mathcal{A}(m_+^2,m_-^2)\cos\bigg{(}\frac{\Delta m t}{2}\bigg{)} + i e^{-i2\beta}\overline{\mathcal{A}}(m_+^2,m_-^2)\sin\bigg{(}\frac{\Delta m t}{2}\bigg{)}\,.
  \label{eq:ampt}
\end{equation}
Here, $\Delta m = m_{B^0_H} - m_{B^0_L}$ is the mass difference between the two physical eigenstates of the $\Bz\text{--}\Bzb$ system,\footnote{
    The notations $\Delta m$ and $\Delta \Gamma$ without subscript are used for brevity, and since there should be no ambiguity over whether the $\Bz\text{--}\Bzb$ or $\Bs\text{--}\Bsb$ system is being referred to.
}
and it has been assumed that there is no \CP violation in mixing so that, in terms of the usual mixing parameters $p$ and $q$ of the $\Bz\text{--}\Bzb$ system, $\frac{q}{p}$ has unit magnitude (see, for example, Refs.~\cite{PDG-review,PDG2022}).
The amplitudes $\mathcal{A}(m_+^2,m_-^2)$ and $\overline{\mathcal{A}}(m_+^2,m_-^2)$ are those for $\Bz \to \Dzb\pip\pim$ and $\Bzb \to \Dz\pip\pim$ decays, respectively, to the position $(m_+^2,m_-^2)$ in the Dalitz plot.
These will be denoted subsequently by $\mathcal{A}_f$ and $\overline{\mathcal{A}}_f$, respectively.
When $\mathcal{A}_f$ and $\overline{\mathcal{A}}_f$ each include contributions with only one set of CKM matrix elements, then there is no \CP\ violation in decay, $\left| \frac{\overline{\mathcal{A}}_f}{\mathcal{A}_f} \right| = 1$, and $\arg\left(\frac{q}{p}\frac{\overline{\mathcal{A}}_f}{\mathcal{A}_f}\right) \equiv \phi_{\rm mix+dec}$. 
For the $\Bz \to D_{\CP}\pip\pim$ case, neglecting contributions from doubly-Cabibbo-suppressed $B$ decay amplitudes, \ie\ from the decay $\Bz \to \Dz\pip\pim$ and the charge-conjugate process, $\phi_{\rm mix+dec} = -2\beta$.  
In Eq.~\eqref{eq:ampt} the weak phase factors in the amplitudes are absorbed into the $e^{-i2\beta}$ term. 
In addition, the width difference in the $\Bz\text{--}\Bzb$ system is taken to be negligible, \ie\ $\Delta\Gamma \approx 0$.
A similar expression to Eq.~\eqref{eq:ampt}, with $\mathcal{A} \leftrightarrow \overline{\mathcal{A}}$ and $-2\beta \leftrightarrow +2\beta$, holds for a neutral $B$ meson tagged as \Bzb\ at time $t=0$.
The symmetry in the Dalitz plot around the line $m_+^2 = m_-^2$ means that for decay to any \CP\ eigenstate $\mathcal{A}(x,y) = \pm \mathcal{A}(y,x)$; the convention that the positive (negative) sign is used for \CP-even (\CP-odd) eigenstates is used in this paper. 
Thus, similar expressions can be used for any \Bz-meson decay to a self-conjugate final state, with the amplitudes expressed as a function of appropriately symmetric co-ordinates.

Returning to the specific $\Bz \to D_{\CP}\pip\pim$ case, Eq.~\eqref{eq:ampt} assumes that the neutral $D$ meson is reconstructed in a \CP-even eigenstate; for \CP-odd $D$ final states, the $+$ sign between the two terms of Eq.~\eqref{eq:ampt} should be replaced by a $-$ sign.
Effects due to charm mixing and \CP\ violation are neglected.  

Squaring Eq.~\eqref{eq:ampt}, and the corresponding equation for a neutral $B$ meson tagged as $\Bzb$ at time $t=0$, expressions for the time-dependent decay rates are obtained:
\begin{multline}
    \Gamma[\Bd \to f(t)] \propto e^{-t/\tau_\Bd}\Big{(}|\mathcal{A}_f|^2+|\overline{\mathcal{A}}_f|^2-2\mathcal{I}m\Big{(}e^{-i2\beta}\mathcal{A}_f^*\overline{\mathcal{A}}_f\Big{)} \sin(\Delta m t)\\+(|\mathcal{A}_f|^2-|\overline{\mathcal{A}}_f|^2)\cos(\Delta m t)\Big{)}\,,
    \label{eq:B0_to_fCP_wrt_time_ideal}
\end{multline}
\vspace{-5ex}
\begin{multline}
    \Gamma[\Bdb \to f(t)] \propto e^{-t/\tau_\Bd}\Big{(}|\mathcal{A}_f|^2+|\overline{\mathcal{A}}_f|^2+2\mathcal{I}m\Big{(}e^{-i2\beta}\mathcal{A}_f^*\overline{\mathcal{A}}_f\Big{)} \sin(\Delta m t)\\-(|\mathcal{A}_f|^2-|\overline{\mathcal{A}}_f|^2)\cos(\Delta m t)\Big{)}\,.
    \label{eq:B0bar_to_fCP_wrt_time_ideal}
\end{multline}
The normalisation is omitted as it is not relevant for the discussion here.
It is assumed that $t$ ranges over non-negative values, as is the case for $B$ mesons produced in LHC collisions.
For $\epem \to \Upsilon(4S) \to B\Bbar$ production both positive and negative values of $t$ are possible, as the tagging information comes from the other decay in the $B\Bbar$ pair; the exponential factor in this case should instead be $e^{-|t|/\tau_\Bd}$.

The coefficient of the sinusoidal term of Eq.~\eqref{eq:B0_to_fCP_wrt_time_ideal} is given by
\begin{equation}
    \label{eq:Im-part}
    -2\mathcal{I}m\Big{(}e^{-i2\beta}\mathcal{A}_f^*\overline{\mathcal{A}}_f\Big{)} = -2\left(\mathcal{I}m(\mathcal{A}_f^*\overline{\mathcal{A}}_f) \,\costb - \mathcal{R}e(\mathcal{A}_f^*\overline{\mathcal{A}}_f) \,\sintb \right) \, .
\end{equation}
In case $\mathcal{A}_f^*\overline{\mathcal{A}}_f$ has no imaginary component, as would be true if the entire $\D_{\CP}\pip\pim$ Dalitz plot were dominated by a single \CP-eigenstate, there is no sensitivity to $\costb$.
It is the interference in the final-state phase-space between different resonances with different \CP compositions --- for example, interference between S- and P-wave $\pip\pim$ components --- that generates the imaginary part of $\mathcal{A}_f^*\overline{\mathcal{A}}_f$ and hence allows \costb\ to be measured.

Without loss of generality, the amplitudes $\mathcal{A}_f$ and $\overline{\mathcal{A}}_f$ can be decomposed into \CP-even ($A_{+\,f}$) and \CP-odd ($A_{-\,f}$) parts,
\begin{equation}
\label{eq:AmplitudeDecomposition}
  \mathcal{A}_f = \frac{A_{+\,f} + A_{-\,f}}{\sqrt{2}}\,, \qquad \overline{\mathcal{A}}_f = \frac{A_{+\,f} - A_{-\,f}}{\sqrt{2}}\,.
\end{equation}
Note that here the \CP\ eigenvalues are those of the decaying neutral $B$ meson rather than of the final state, though in the absence of \CP\ violation in decay, as assumed here, the two are equivalent.
The impact of non-negligible \CP\ violation in decay will be considered in Sec.~\ref{sec:CPVinDecay}.
The previously mentioned phase convention implies that the $B$ meson \CP\ eigenstates are given by $\left\vert B_\pm \right\rangle = \frac{1}{\sqrt{2}}\left( \left\vert \Bz \right\rangle \pm \left\vert \Bzb \right\rangle \right)$, which is used in Eq.~\eqref{eq:AmplitudeDecomposition} but does not affect the final results. 

In the $D_{\CP}\pi^+\pi^-$ final state with a \CP-even $D$ meson, since the $\pi^+\pi^-$ system must also be \CP-even, the overall \CP\ eigenvalue is equal to $(-1)^l$ where $l$ is the angular momentum between the $D$ meson and the $\pi^+\pi^-$ system.
By angular momentum conservation in the $B$-meson decay, $l$ is also the angular momentum in the $\pi^+\pi^-$ system.
Hence, in the absence of \CP\ violation in decay, $A_{+\,f}$ ($A_{-\,f}$) corresponds to amplitudes where the $\pi^+\pi^-$ system is in an even (odd) partial wave.
This means that $A_{+\,f}$ ($A_{-\,f}$) is symmetric (antisymmetric) about the Dalitz-plot symmetry line $m_+^2 = m_-^2$.
Since any amplitude can be decomposed into a sum of partial waves, with the sum potentially running to infinity, this formalism includes not only amplitudes corresponding to $\pi\pi$ resonances, but also those corresponding to $D\pi$ resonances.

\begin{figure}[tb]
    \centering
    \includegraphics[width=0.4\linewidth]{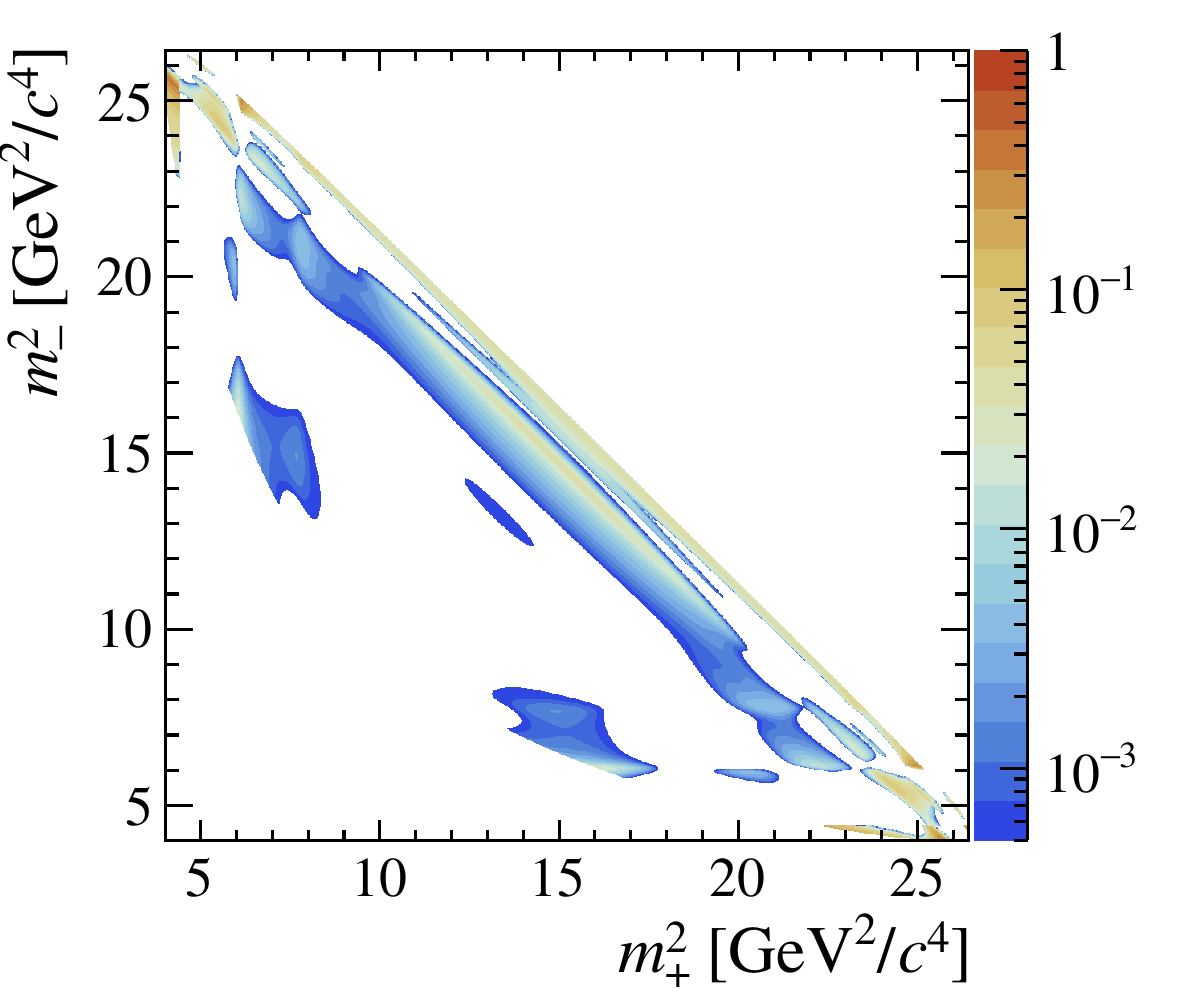}
    \includegraphics[width=0.4\linewidth]{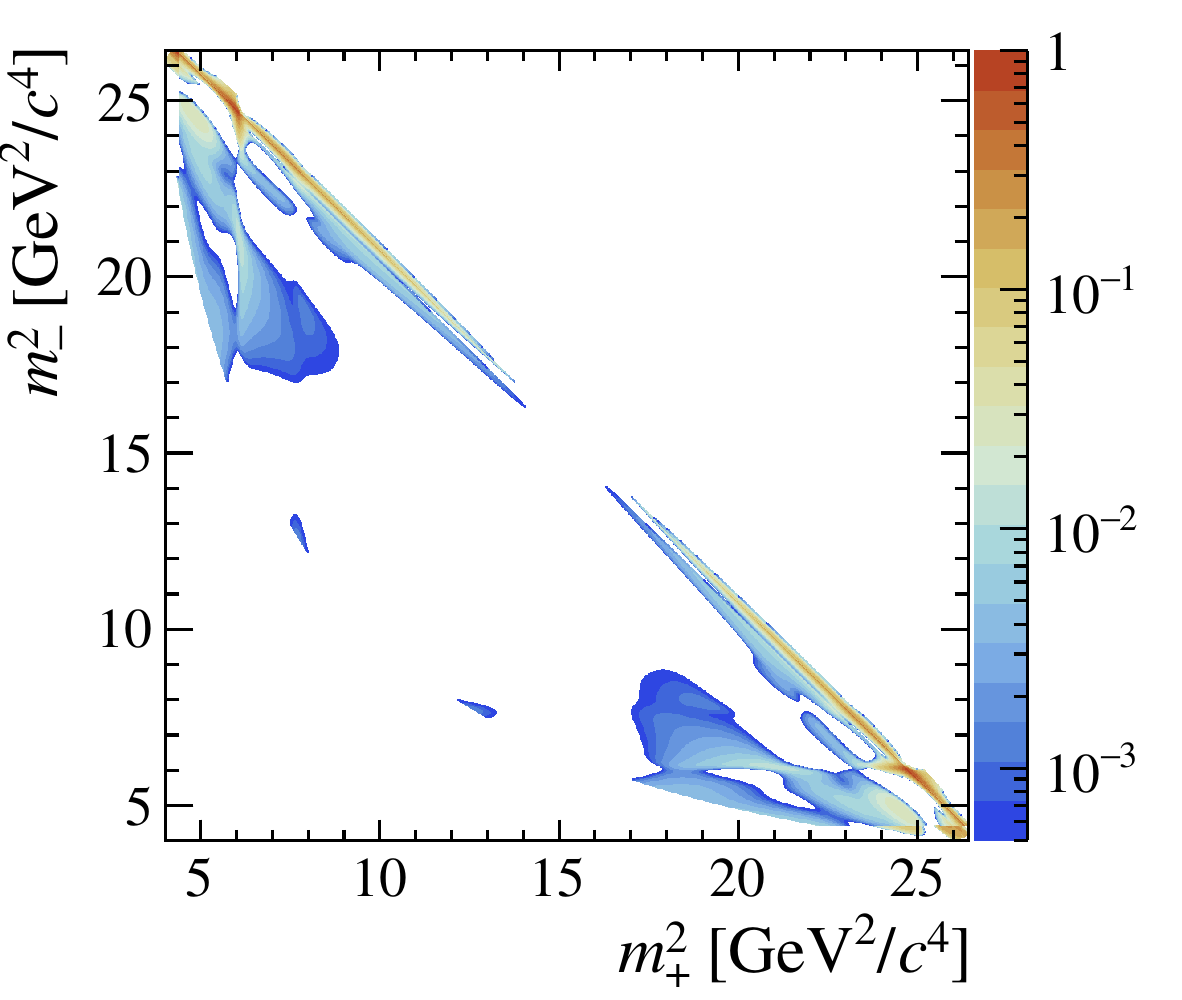}\\
    \includegraphics[width=0.4\linewidth]{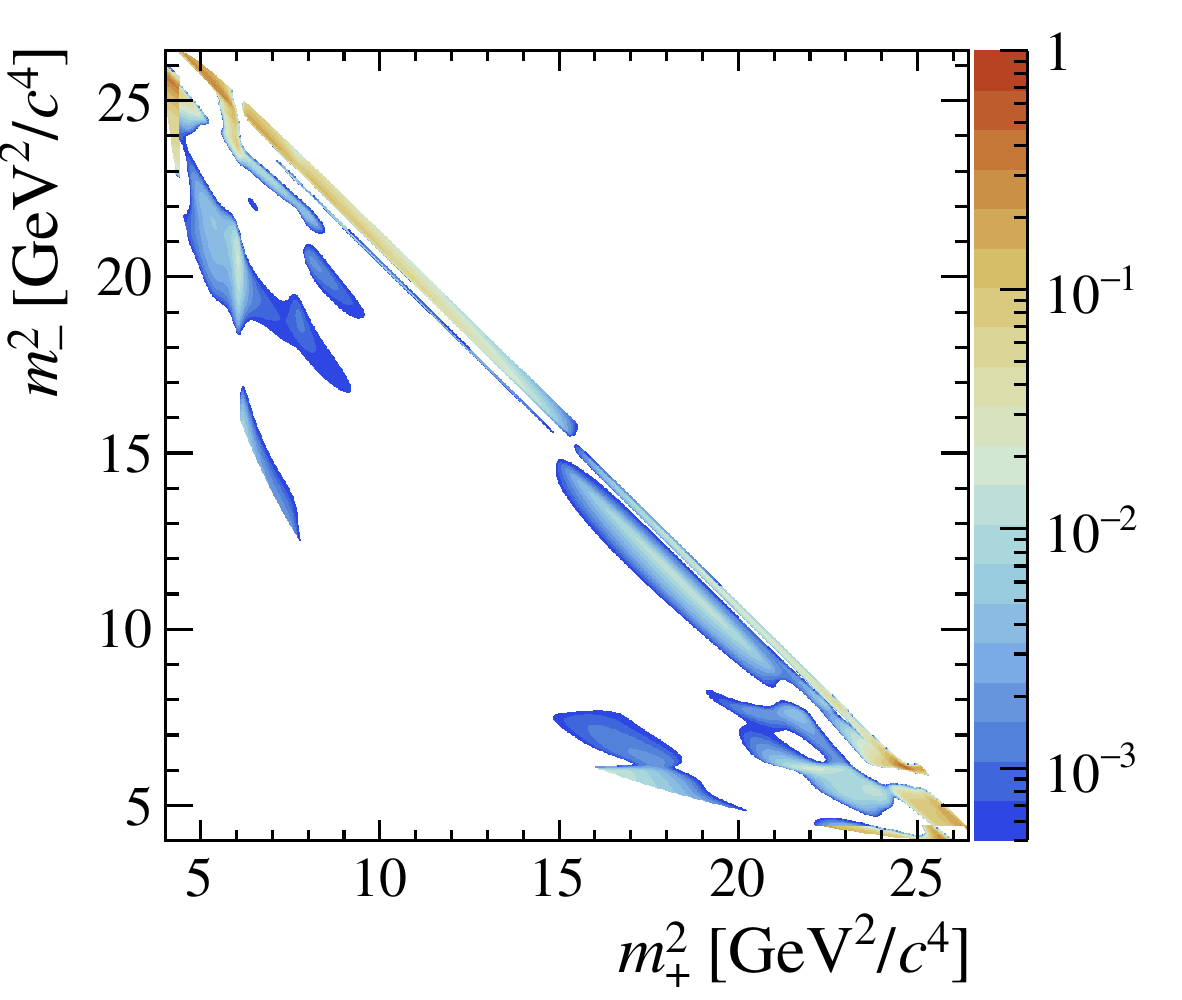}
    \includegraphics[width=0.4\linewidth]{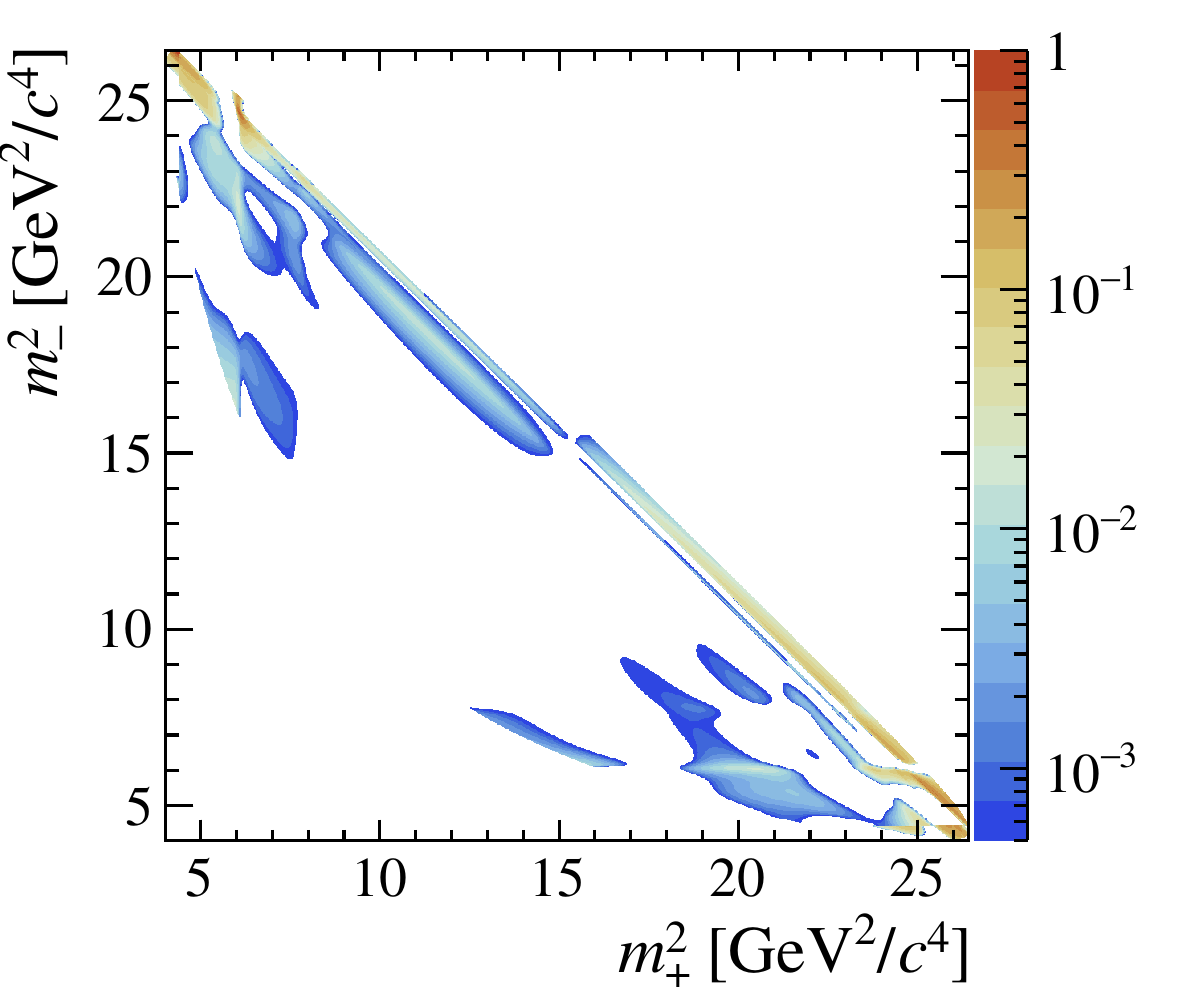}
    \caption{
        Dalitz-plot distributions of (top left) $\mathcal{R}e(\mathcal{A}_f^*\overline{\mathcal{A}}_f)$ for $\mathcal{R}e(\mathcal{A}_f^*\overline{\mathcal{A}}_f)>0$, (top right) $-\mathcal{R}e(\mathcal{A}_f^*\overline{\mathcal{A}}_f)$ for $\mathcal{R}e(\mathcal{A}_f^*\overline{\mathcal{A}}_f)<0$, (bottom left) $\mathcal{I}m(\mathcal{A}_f^*\overline{\mathcal{A}}_f)$ for $\mathcal{I}m(\mathcal{A}_f^*\overline{\mathcal{A}}_f)>0$, and (bottom right) $-\mathcal{I}m(\mathcal{A}_f^*\overline{\mathcal{A}}_f)$ for $\mathcal{I}m(\mathcal{A}_f^*\overline{\mathcal{A}}_f)<0$, with the Dalitz-plot model for $B^0 \to D_{\CP} \pip\pim$ decays from Ref.~\cite{LHCb-PAPER-2014-070}, illustrating the symmetry (antisymmetry) of $\mathcal{R}e(\mathcal{A}_f^*\overline{\mathcal{A}}_f)$ ($\mathcal{I}m(\mathcal{A}_f^*\overline{\mathcal{A}}_f)$).
        The $z$-axis scale is arbitrary, but common for the four plots.
    }
    \label{fig:re_im_illustration}
\end{figure}

The interference term between the \Bz\ and \Bzb\ decay amplitudes can therefore be written as 
\begin{eqnarray}
  \mathcal{A}_f^*\overline{\mathcal{A}}_f & = & \frac{1}{2}\left( A^*_{+\,f} + A^*_{-\,f} \right)\left( A_{+\,f} - A_{-\,f} \right)\,, \\
   & = & \frac{1}{2}\left\{ \left|A_{+\,f}\right|^2-\left|A_{-\,f}\right|^2 + 2 i \mathcal{I}m\left(A^*_{-\,f}A_{+\,f}\right)\right\}\,,
\end{eqnarray}
and thus
\begin{eqnarray}
  \mathcal{R}e(\mathcal{A}_f^*\overline{\mathcal{A}}_f) & = & \frac{\left|A_{+\,f}\right|^2-\left|A_{-\,f}\right|^2}{2}\,, \label{eq:symRel-Re} \\
  \mathcal{I}m(\mathcal{A}_f^*\overline{\mathcal{A}}_f) & = & \mathcal{I}m\left(A^*_{-\,f}A_{+\,f}\right) \,. \label{eq:symRel-Im}
\end{eqnarray}
It is clear that $\mathcal{R}e(\mathcal{A}_f^*\overline{\mathcal{A}}_f)$ ($\mathcal{I}m(\mathcal{A}_f^*\overline{\mathcal{A}}_f)$) is symmetric (antisymmetric) about the Dalitz-plot symmetry line (\ie\ under interchange of $m_+^2$ and $m_-^2$).
This is illustrated in Fig.~\ref{fig:re_im_illustration} using the model for $\Bz \to D_{\CP}\pip\pim$ decays described in the next section.
Thus, from Eq.~\eqref{eq:Im-part} it can be seen that a simple integration over the Dalitz plot will retain sensitivity to \sintb, through the $\mathcal{R}e(\mathcal{A}_f^*\overline{\mathcal{A}}_f)$ term, though diluted by cancellation of \CP-even and \CP-odd contributions.
The sensitivity to \costb\ will be completely cancelled out.

Weighting functions, which can be applied to the data to enable visualisation of the sensitivity to \sintb\ and \costb, can be proposed with the following logic.
In order to visualise \sintb\ (\costb), the function should be symmetric (antisymmetric) about the symmetry line, so that the relevant terms are retained when integrating over the Dalitz plot.
The function should also carry the same sign as $\mathcal{R}e(\mathcal{A}_f^*\overline{\mathcal{A}}_f)$ ($\mathcal{I}m(\mathcal{A}_f^*\overline{\mathcal{A}}_f)$) to avoid cancellation of Dalitz-plot regions where these terms have opposite signs.
Finally, in the limit of a pure \CP-eigenstate amplitude, the weighting function should be constant across the Dalitz plot, as in this case any variation in the weighting would dilute the asymmetry.
Functions that satisfy these criteria, for the visualisation of the sensitivity to \sintb\ and \costb, are 
\begin{equation}
\label{eq:weighting-functions}
    w^{\mathcal{R}e}\!\left(m_+^2,m_-^2\right) = \frac{2\,\mathcal{R}e(\mathcal{A}_f^*\overline{\mathcal{A}}_f)}{|\mathcal{A}_f|^2+|\overline{\mathcal{A}}_f|^2}\,,
    \quad\text{and}\quad
    w^{\mathcal{I}m}\!\left(m_+^2,m_-^2\right) = \frac{2\,\mathcal{I}m(\mathcal{A}_f^*\overline{\mathcal{A}}_f)}{|\mathcal{A}_f|^2+|\overline{\mathcal{A}}_f|^2}\,,
\end{equation}
respectively.
The normalisation is chosen such that these functions are each in the range $\left[-1,+1\right]$, but this does not affect the outcome.

The decay-time asymmetry is defined in terms of the time-dependent decay rates of Eqs.~\eqref{eq:B0_to_fCP_wrt_time_ideal} and~\eqref{eq:B0bar_to_fCP_wrt_time_ideal},
\begin{equation}
\label{eq:simple-acpt}
    A_{\CP}(t) \equiv \frac{\Gamma[\Bdb \to f(t)]-\Gamma[\Bd \to f(t)]}{\Gamma[\Bdb \to f(t)]+\Gamma[\Bd \to f(t)]}\,.
\end{equation}
When forming weighted decay-time asymmetry plots, each $\Gamma$ term is integrated over the Dalitz plot.
(It is also integrated over the range of each decay-time bin, but that is not relevant to the discussion here.)
The impact of the weights on all terms of Eqs.~\eqref{eq:B0_to_fCP_wrt_time_ideal} and~\eqref{eq:B0bar_to_fCP_wrt_time_ideal} must therefore be considered.
It can be seen that 
\begin{equation}
    |\mathcal{A}_f|^2+|\overline{\mathcal{A}}_f|^2 = \left|A_{+\,f}\right|^2+\left|A_{-\,f}\right|^2\,,
\end{equation}
is symmetric about the Dalitz-plot symmetry line, while
\begin{equation}
    |\mathcal{A}_f|^2-|\overline{\mathcal{A}}_f|^2 = 2\,\mathcal{R}e\left(A^*_{-\,f}A_{+\,f}\right)
\end{equation}
is antisymmetric.
Therefore, at least in the case of an antisymmetric weighting function, it is better not to weight the expressions in the denominator of the asymmetry to avoid a potential ``divide by zero'' problem.

Weighted decay-time asymmetries are hence defined as 
\begin{equation}
\label{eq:weighted-acpt}
    A^{w-\mathcal{R}e/\mathcal{I}m}_{\CP}(t) \equiv \frac{\int_{\rm DP} w^{\mathcal{R}e/\mathcal{I}m}\!\left(m_+^2,m_-^2\right) \left( \Gamma[\Bdb \to f(t)] - \Gamma[\Bd \to f(t)]\right) {\rm d}\Omega}{\int_{\rm DP} \left( \Gamma[\Bdb \to f(t)]+\Gamma[\Bd \to f(t)] \right) {\rm d}\Omega}\,.
\end{equation}
where $w^{\mathcal{R}e/\mathcal{I}m}$ is one of the weighting functions from Eq.~\eqref{eq:weighting-functions} and ${\rm d}\Omega$ represents an element in the Dalitz-plot phase-space.
For the first weighting function, the weighted asymmetry reduces to 
\begin{eqnarray}
\label{eq:wacp_re_def}
  A^{w-\mathcal{R}e}_{\CP}(t) & = & S^{w-\mathcal{R}e} \sin(\Delta m t)\,, \\ %% - C^{w-\mathcal{R}e} \cos(\Delta m t) -- cosine term disappears!
\label{eq:wS_re_def}
  \text{where} \ S^{w-\mathcal{R}e} & = & \sintb \left(\frac{ -4 \int_{\rm DP} \left[ \mathcal{R}e(\mathcal{A}_f^*\overline{\mathcal{A}}_f) \right]^2/\left(|\mathcal{A}_f|^2+|\overline{\mathcal{A}}_f|^2\right) {\rm d}\Omega }{\int_{\rm DP} \left(|\mathcal{A}_f|^2+|\overline{\mathcal{A}}_f|^2\right) {\rm d}\Omega}\right)\,,
\end{eqnarray}
while for the second weighting function instead 
\begin{eqnarray}
\label{eq:wacp_im_def}
  A^{w-\mathcal{I}m}_{\CP}(t) & = & S^{w-\mathcal{I}m} \sin(\Delta m t) - C^{w-\mathcal{I}m} \cos(\Delta m t)\,,\\
\label{eq:wS_im_def}
  \text{where} \ S^{w-\mathcal{I}m} & = & \costb \left( \frac{ 4 \int_{\rm DP} \left[ \mathcal{I}m(\mathcal{A}_f^*\overline{\mathcal{A}}_f) \right]^2/\left(|\mathcal{A}_f|^2+|\overline{\mathcal{A}}_f|^2\right) {\rm d}\Omega }{\int_{\rm DP} \left(|\mathcal{A}_f|^2+|\overline{\mathcal{A}}_f|^2\right) {\rm d}\Omega} \right) \,,\\
\label{eq:wC_im_def}
  \text{and} \ C^{w-\mathcal{I}m} & = & \frac{ 2 \int_{\rm DP} \left(|\mathcal{A}_f|^2-|\overline{\mathcal{A}}_f|^2\right) \mathcal{I}m(\mathcal{A}_f^*\overline{\mathcal{A}}_f)/\left(|\mathcal{A}_f|^2+|\overline{\mathcal{A}}_f|^2\right) {\rm d}\Omega}{\int_{\rm DP} \left(|\mathcal{A}_f|^2+|\overline{\mathcal{A}}_f|^2\right) {\rm d}\Omega}\,.
\end{eqnarray}
 
Hence, $A^{w-\mathcal{R}e}_{\CP}(t)$ and $A^{w-\mathcal{I}m}_{\CP}(t)$ can be used to visualise the determination of \sintb\ and \costb, respectively, both appearing as sine waves.
The asymmetry $A^{w-\mathcal{I}m}_{\CP}(t)$ will also show a cosine oscillation, with magnitude independent of weak phase factors.
The magnitudes of the oscillations are modulated by hadronic factors, but these constants are calculable assuming that the Dalitz-plot model is known.
In the pure \CP-eigenstate limit, the hadronic factor in Eq.~\eqref{eq:wS_re_def} is equal to $-1$, while those in Eqs.~\eqref{eq:wS_im_def} and~\eqref{eq:wC_im_def} are equal to zero.

\section{Illustration of the method}
\label{sec:illustration}

To illustrate the method, it is necessary to specify the amplitudes $\mathcal{A}_f$ and $\overline{\mathcal{A}}_f$.
As previously, the $\Bz \to D_{\CP}\pip\pim$ decay is taken as an example, with amplitude model taken from the analysis by the LHCb collaboration~\cite{LHCb-PAPER-2014-070}, and implemented in the {\sc Laura}$^{++}$ software package~\cite{Back:2017zqt}.
Specifically, the results of the isobar model from Ref.~\cite{LHCb-PAPER-2014-070} are used.
The largest components in the model are $D\rho^0$, $D(\pip\pim)_{\rm S\,wave}$ (including $Df_0(500)$) and $Df_2(1270)$ together with $D_2^*(2460)^-\pip$, $D_0^*(2400)^-\pip$ and a $(D\pim)_{\rm P\,wave}\pip$ term.
The $D^*(2010)^-\pip$ component is vetoed with the requirement $m(D\pipm) > 2.1 \gevcc$, since it is too narrow to contribute significant interference effects that would enhance the sensitivity to \sintb\ and \costb.

The hadronic factors in Eqs.~\eqref{eq:wS_re_def},~\eqref{eq:wS_im_def} and~\eqref{eq:wC_im_def} are evaluated numerically with this model, and are found to be 
\begin{eqnarray}
\label{eq:wfactor1}
\frac{ -4 \int_{\rm DP} \left[ \mathcal{R}e(\mathcal{A}_f^*\overline{\mathcal{A}}_f) \right]^2/\left(|\mathcal{A}_f|^2+|\overline{\mathcal{A}}_f|^2\right) {\rm d}\Omega }{\int_{\rm DP} \left(|\mathcal{A}_f|^2+|\overline{\mathcal{A}}_f|^2\right) {\rm d}\Omega} & = & -0.267\,, \\ %-0.187 [The commented value includes a factor of sin(2beta).]
\label{eq:wfactor2}
\frac{ 4 \int_{\rm DP} \left[ \mathcal{I}m(\mathcal{A}_f^*\overline{\mathcal{A}}_f) \right]^2/\left(|\mathcal{A}_f|^2+|\overline{\mathcal{A}}_f|^2\right) {\rm d}\Omega }{\int_{\rm DP} \left(|\mathcal{A}_f|^2+|\overline{\mathcal{A}}_f|^2\right) {\rm d}\Omega} & = & \phantom{-}0.229\,,\\ %0.164 [The commented value includes a factor of cos(2beta).]
\label{eq:wfactor3}
\frac{2 \int_{\rm DP} \left(|\mathcal{A}_f|^2-|\overline{\mathcal{A}}_f|^2\right) \mathcal{I}m(\mathcal{A}_f^*\overline{\mathcal{A}}_f)/\left(|\mathcal{A}_f|^2+|\overline{\mathcal{A}}_f|^2\right) {\rm d}\Omega}{\int_{\rm DP} \left(|\mathcal{A}_f|^2+|\overline{\mathcal{A}}_f|^2\right) {\rm d}\Omega} & = & -0.002\,.
\end{eqnarray}

\begin{figure}[!tb]
    \centering
    \includegraphics[width=0.6\linewidth]{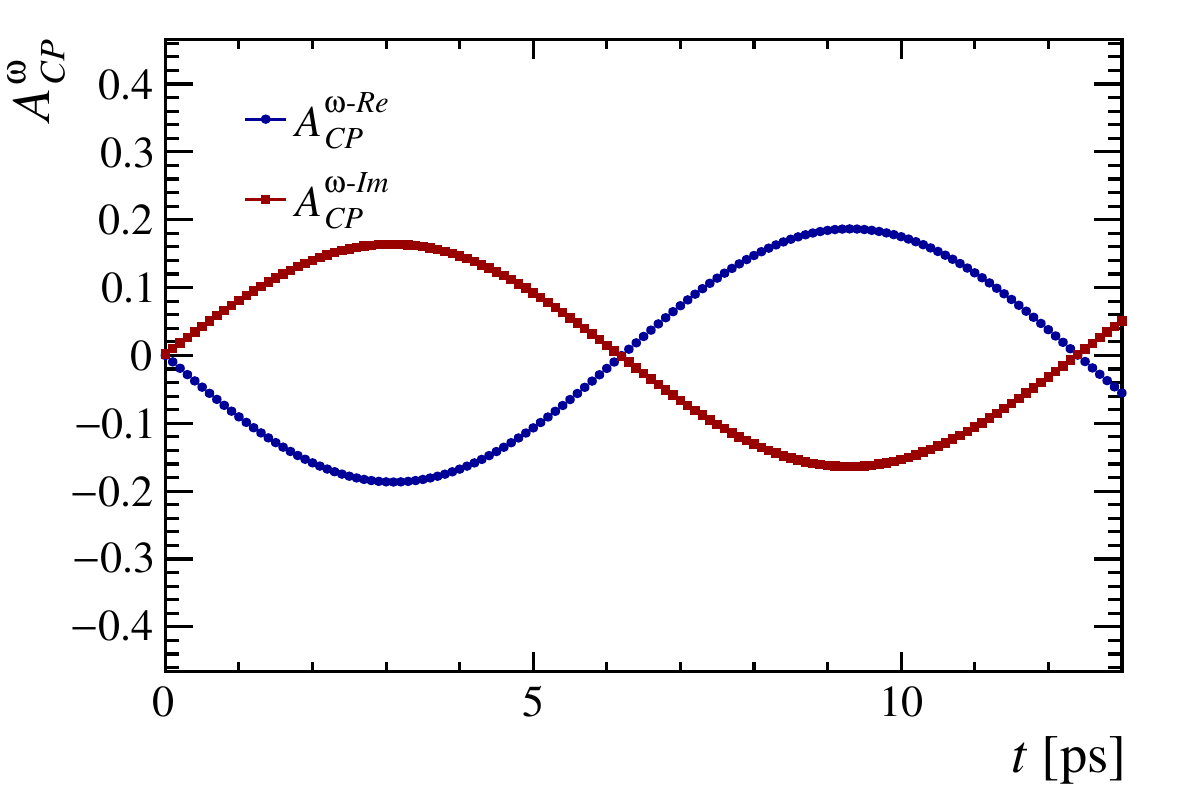}
    \caption{Decay-time dependence of $A^{w-\mathcal{R}e}_{\CP}(t)$ and $A^{w-\mathcal{I}m}_{\CP}(t)$ in $\Bz \to D_{\CP}\pip\pim$ decays with the Dalitz-plot model from Ref.~\cite{LHCb-PAPER-2014-070}.}
    \label{fig:wacp_reim}
\end{figure}

Distributions of $A^{w-\mathcal{R}e}_{\CP}(t)$ and $A^{w-\mathcal{I}m}_{\CP}(t)$, as defined in Eqs.~\eqref{eq:wacp_re_def} and~\eqref{eq:wacp_im_def}, obtained with this model are shown in Fig.~\ref{fig:wacp_reim}.
Here, \sintb\ and \costb\ are set at their world average values (corresponding to $\beta = 22.2\degrees$~\cite{HFLAV21}, so 0.700 and 0.714, respectively).
These correspond to the distributions that would be obtained in the limit of infinite statistics with no experimental effects such as imperfect flavour tagging.

Samples of pseudoexperiments are generated in order to approximate what the weighted decay-time asymmetries might look like in an experiment with currently plausible precision.
The $\Bz \to D\pip\pim$ yield obtained by the LHCb collaboration in Ref.~\cite{LHCb-PAPER-2014-070}, of $\sim10\,000$, is scaled by 0.14 to account for the difference in branching fractions between the $D \to \Kp\pim$ channel used in that analysis and the combined $D \to \Kp\Km$ and $D \to \pip\pim$ decays that are expected to provide the largest $D_{\CP}$ samples.
Further scaling factors, indicated in parentheses, are applied to account for: the difference in sample size available to the LHCb collaboration in Ref.~\cite{LHCb-PAPER-2014-070} to the total recorded in the LHC Run~1 and 2 operation periods (5); the facts that in Ref.~\cite{LHCb-PAPER-2014-070} only a subset of possible trigger lines was used (2) and that very tight selection requirements were imposed (1.5); the typical LHCb flavour tagging power on similar decays (0.08)~\cite{LHCb-PAPER-2011-027,LHCb-PAPER-2015-027,LHCB-PAPER-2016-037,LHCb-PAPER-2016-039,LHCb-PAPER-2019-036}.
Rounding up optimistically, an estimate of 2000 is obtained for the equivalent yield of perfectly tagged $\Bz \to D_{\CP}\pip\pim$ decays available in the Run~1 and~2 LHCb data sample.

\begin{figure}[!tb]
    \centering
    \includegraphics[width=0.48\linewidth]{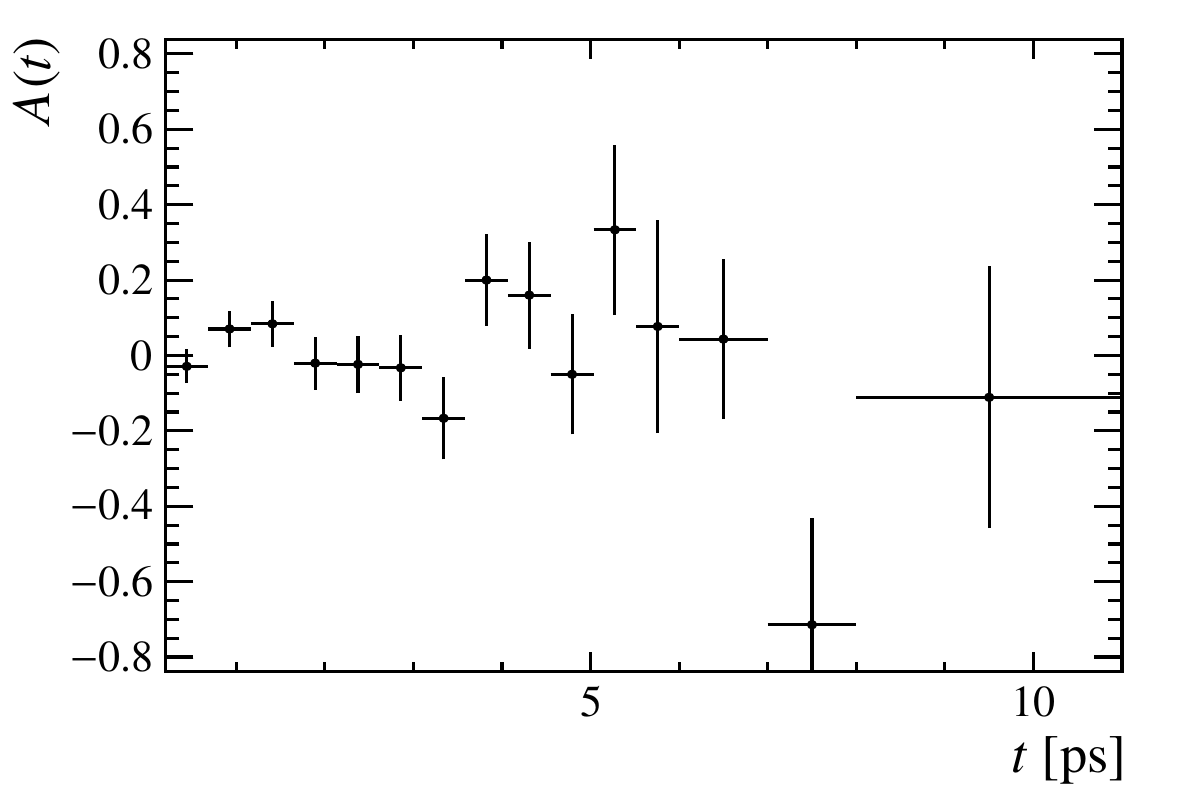} \\
    \includegraphics[width=0.48\linewidth]{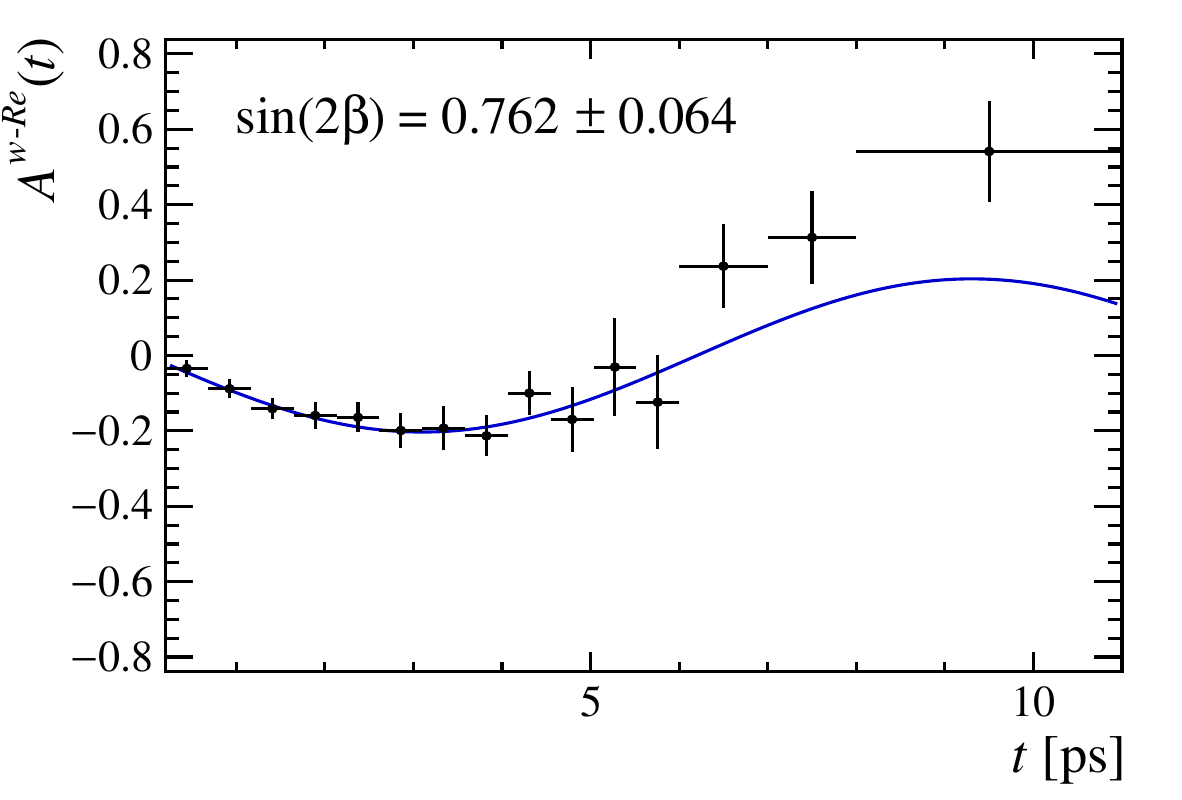}
    \includegraphics[width=0.48\linewidth]{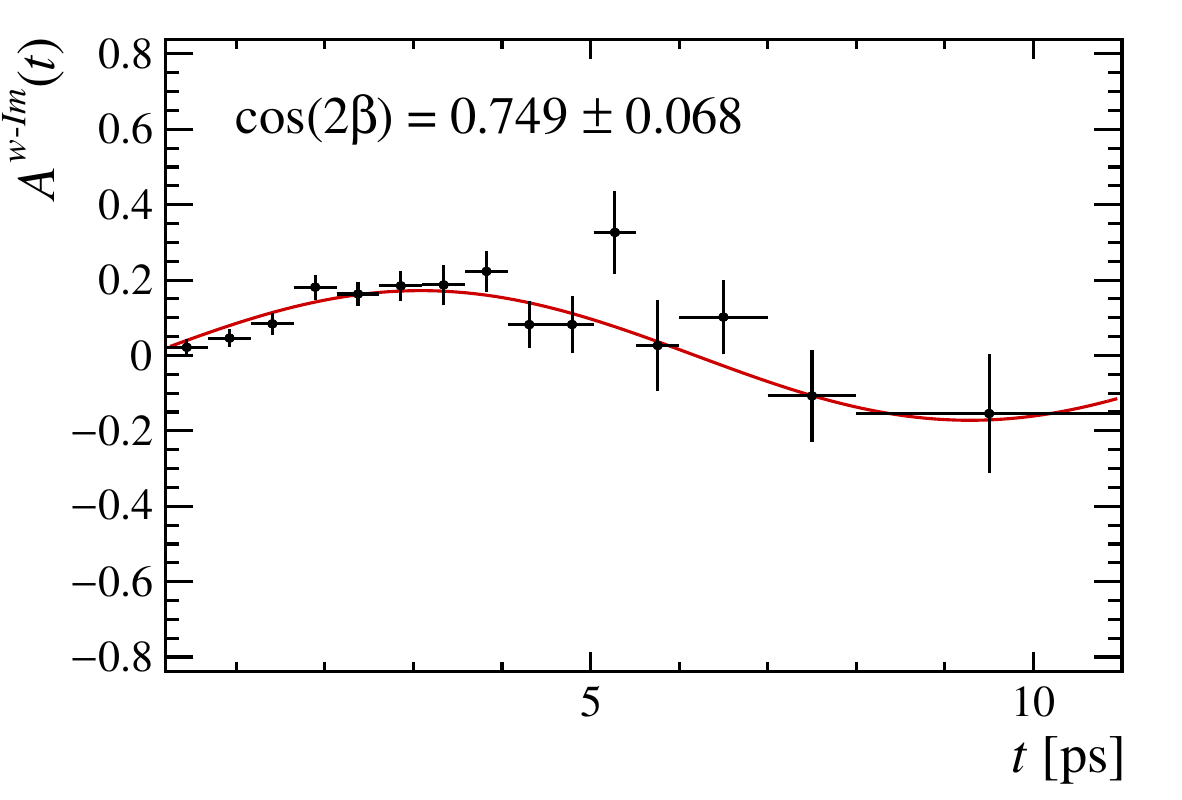} 
    \caption{Distributions of time-dependent asymmetry with the time-dependent decay rates in the numerator (top)~unweighted, (bottom left)~weighted by $w^{\mathcal{R}e}$, and (bottom right)~weighted by $w^{\mathcal{I}m}$.
    These are obtained from pseudoexperiments generated with the Dalitz-plot model from Ref.~\cite{LHCb-PAPER-2014-070}, and with sample size corresponding to the equivalent perfectly tagged yield expected to be available in the LHCb data sample.
    The statistical uncertainties are evaluated with a bootstrap method~\cite{efron:1979}. 
    Results of the fits described in the text are also shown.}
    \label{fig:wacp_toy}
\end{figure}

Neglecting other experimental effects, such as background, efficiency variations across the Dalitz plot or with decay time, and resolution, samples of this size are generated and then weighted with the functions of Eq.~\eqref{eq:weighting-functions}.
Summing over the range of each decay-time bin and the whole Dalitz plot, the weighted decay-time asymmetries shown in Fig.~\ref{fig:wacp_toy} are constructed.
The unweighted decay-time asymmetry is also shown for comparison.
The utility of the weighting functions in visualising the physics effects of interest is evident.

It may be noted that, once these weighted decay-time asymmetries have been formed, values of \sintb\ and \costb\ can be obtained by simple one-dimensional fits to them, assuming that the hadronic factors are fixed.
Such fits would not be expected to give values as precise as can be obtained from the full unbinned decay-time-dependent amplitude analysis, but could be used to obtain a cross-check of the result.
The values of \sintb\ and \costb\ obtained by fitting the weighted asymmetries, shown in Fig.~\ref{fig:wacp_toy}, are $\sintb = 0.762 \pm 0.064$ and $\costb = 0.749 \pm 0.068$; for comparison the values obtained from the full decay-time-dependent amplitude fit to these samples are $\sintb = 0.742 \pm 0.045$ and $\costb = 0.735 \pm 0.045$.

Alternatively, if the values of \sintb\ and \costb\ are fixed, the weighted decay-time asymmetries can be used to check that the values of the hadronic factors in data are consistent with those obtained from the model.
In either case, the method provides a useful new way to visualise the effects being measured, and to check the consistency of the data with the fit result.  

\section{Experimental effects}
\label{sec:experiment}

The decay-time distributions measured in experiment are not given only by the expressions of Eqs.~\eqref{eq:B0_to_fCP_wrt_time_ideal} and~\eqref{eq:B0bar_to_fCP_wrt_time_ideal}, but are modified by contributions from background processes, asymmetries in the $\Bz$--$\Bzb$ production rate or detection probability, efficiency variations across the decay phase-space or with decay time, smearing of the distributions due to resolution, and imperfect flavour tagging.  
For most of these it is possible to apply corrections to the data, as discussed below, so that the asymmetry curve still represents that of the underlying physics.  
This is, however, not essential -- the asymmetries measured in data without further corrections would still be useful to visualise the sensitivity to \sintb\ and \costb, although the impact of the experimental effects would need to be assessed on a case-by-case basis.

Background can be subtracted statistically using one of a number of methods.
The \sPlot\ method and its variants~\cite{Pivk:2004ty,Dembinski:2021kim}, based on the application of a signal weight for each candidate in the selected sample, provide an attractive option for many decays.
Weights can also be used to correct for efficiency variation across the phase-space, as has been done in numerous Dalitz-plot analyses (see, \eg, Ref.~\cite{LHCb-PAPER-2012-018}), and to correct for detection asymmetries, assuming that the size of these effects is known from appropriately calibrated simulation.   
Variation with decay-time of the efficiency will not affect the asymmetry as long as the bin size is chosen such that variations within each bin are small; if this is not the case, corrections can be applied.
Similarly, effects due to imperfect flavour tagging can be corrected for either on a candidate-by-candidate basis or collectively, as the size of the required correction will be known from data control samples.
An example of a decay-time asymmetry being visualised with candidate-by-candidate flavour-tagging corrections applied can be found in Ref.~\cite{LHCb-PAPER-2023-013}.
A non-zero asymmetry in the production rates of \Bz\ and \Bzb\ mesons will introduce a decay-time-independent offset, that can also be corrected for assuming that the size of the effect is known.  

The only remaining effects to consider are those due to resolution.
Non-negligible effects of resolution in the decay-time variable itself will result in dilution of the oscillations, that must be accounted for when interpreting the visualisation of the data.
This is the case for the asymmetries obtained by the BaBar and Belle experiments, including those shown in Fig.~\ref{fig:sin2beta-Bfactories}.
In the LHCb experiment the decay-time resolution is typically negligible compared to the period of $\Bz$--$\Bzb$ oscillations, but cannot be overlooked when considering $\Bs$--$\Bsb$ oscillations, as for example in Refs.~\cite{LHCb-PAPER-2023-016,LHCb-PAPER-2017-008,LHCb-PAPER-2021-005}.

Smearing of the determination of the position in phase-space of each $B$ decay would mean that the weight applied to each candidate will be calculated at the measured position, rather than the true position.  
As such, this could have non-negligible effects on the weighting procedure if the resolution is not much smaller than the scale on which the amplitudes $\mathcal{A}_f$ and $\overline{\mathcal{A}}_f$ are varying, which for a Dalitz plot can be characterised by the width of the narrowest contributing resonance.  
The ability to impose kinematic constraints when determining the phase-space position will often mean this is not a concern.
In cases where the effect must be accounted for, it will result in a modification of the hadronic factors of Eqs.~\eqref{eq:wS_re_def},~\eqref{eq:wS_im_def} and~\eqref{eq:wC_im_def} that will be calculable for a known amplitude model and resolution function.

\section{Non-zero width difference}
\label{sec:DeltaGamma}

To be applied to \Bs\ decays, the effect of non-zero decay width difference must be accounted for.
This results in modifications to Eqs.~\eqref{eq:B0_to_fCP_wrt_time_ideal} and~\eqref{eq:B0bar_to_fCP_wrt_time_ideal}: in addition to replacing $\Bz \to \Bs$, $\Bzb \to \Bsb$ and $2\beta \to -2\beta_s$,\footnote{
    The sign-flip is due to a convention in the definition of $\beta_s$, to make the value positive in the Standard Model~\cite{PDG-review}.
} the $|\mathcal{A}_f|^2+|\overline{\mathcal{A}}_f|^2$ term is multiplied by $\cosh(\Delta\Gamma t/2)$, and an additional term $-2\mathcal{R}e\Big{(}e^{i2\beta_s}\mathcal{A}_f^*\overline{\mathcal{A}}_f\Big{)} \sinh(\Delta\Gamma t/2)$ contributes to both \Bs\ and \Bsb\ decay rates.\footnote{
    The convention where $\Delta \Gamma$ is positive in the $\Bs$--$\Bsb$ system is used~\cite{PDG-review}.
}
Since this term appears with the same sign in both rates it does not contribute to the numerator of the asymmetry, and since the denominator is unweighted the replacements for Eqs.~\eqref{eq:wacp_re_def}--\eqref{eq:wC_im_def} are
\small 
\begin{eqnarray}
\label{eq:wacp_re_def-Bs}
  A^{w-\mathcal{R}e}_{\CP}(t) & = & S^{w-\mathcal{R}e} \sin(\Delta m t)\,, \\ %% - C^{w-\mathcal{R}e} \cos(\Delta m t) -- cosine term disappears!
\label{eq:wS_re_def-Bs}
  \!\!\!\!\!\text{where}\ S^{w-\mathcal{R}e} & = & \frac{ -4\sintbs \int_{\rm DP}\left[ \mathcal{R}e(\mathcal{A}_f^*\overline{\mathcal{A}}_f) \right]^2/\left(|\mathcal{A}_f|^2+|\overline{\mathcal{A}}_f|^2\right) {\rm d}\Omega }{\int_{\rm DP}\left(|\mathcal{A}_f|^2+|\overline{\mathcal{A}}_f|^2\right)\cosh(\Delta\Gamma t/2) -2\mathcal{R}e\Big{(}e^{i2\beta_s}\mathcal{A}_f^*\overline{\mathcal{A}}_f\Big{)} \sinh(\Delta\Gamma t/2)\,{\rm d}\Omega} \,,\\
\label{eq:wacp_im_def-Bs}
  A^{w-\mathcal{I}m}_{\CP}(t) & = & S^{w-\mathcal{I}m} \sin(\Delta m t) - C^{w-\mathcal{I}m} \cos(\Delta m t)\,,\\
\label{eq:wS_im_def-Bs}
  \!\!\!\!\!\text{where}\ S^{w-\mathcal{I}m} & = & \frac{ 4\costbs \int_{\rm DP}\left[ \mathcal{I}m(\mathcal{A}_f^*\overline{\mathcal{A}}_f) \right]^2/\left(|\mathcal{A}_f|^2+|\overline{\mathcal{A}}_f|^2\right) {\rm d}\Omega }{\int_{\rm DP}\left(|\mathcal{A}_f|^2+|\overline{\mathcal{A}}_f|^2\right)\cosh(\Delta\Gamma t/2) -2\mathcal{R}e\Big{(}e^{i2\beta_s}\mathcal{A}_f^*\overline{\mathcal{A}}_f\Big{)} \sinh(\Delta\Gamma t/2)\,{\rm d}\Omega} \,,\\
\label{eq:wC_im_def-Bs}
  \!\!\!\!\!\text{and}\ C^{w-\mathcal{I}m} & = & \frac{ 2 \int_{\rm DP}\left(|\mathcal{A}_f|^2-|\overline{\mathcal{A}}_f|^2\right) \mathcal{I}m(\mathcal{A}_f^*\overline{\mathcal{A}}_f)/\left(|\mathcal{A}_f|^2+|\overline{\mathcal{A}}_f|^2\right) {\rm d}\Omega}{\int_{\rm DP}\left(|\mathcal{A}_f|^2+|\overline{\mathcal{A}}_f|^2\right)\cosh(\Delta\Gamma t/2) -2\mathcal{R}e\Big{(}e^{i2\beta_s}\mathcal{A}_f^*\overline{\mathcal{A}}_f\Big{)} \sinh(\Delta\Gamma t/2)\,{\rm d}\Omega}\,.
\end{eqnarray}
\normalsize

In practice, the modification of the asymmetries due to the non-zero value of $\Delta\Gamma$ will make little difference.
The value of $\Delta\Gamma/\Gamma$ in the $\Bs$--$\Bsb$ system, $0.126 \pm 0.007$~\cite{HFLAV21} means that after five \Bs\ lifetimes the values of $\left(\cosh(\Delta\Gamma t/2), \sinh(\Delta\Gamma t/2)\right)$ are $(1.05,0.32)$ --- a relatively modest change from the values $(1,0)$ at $t=0$.
At very large numbers of lifetimes the extra term can significantly enhance or suppress the oscillations, depending on its sign, but there will be little data available to observe this effect.
Thus it has become common in experimental analyses of \Bs\ decays to show oscillations folded by the period $2\pi/(\Delta m)$ (see, for example, Refs.~\cite{LHCb-PAPER-2023-016,LHCb-PAPER-2017-008,ATLAS:2020lbz,CMS:2020efq}), ignoring the modulation caused by the extra term at high decay times.  
This approach can also be used for the weighted asymmetries.

Nonetheless, the extra terms involving $\Delta\Gamma$ can have a noticeable effect in the untagged decay-time distribution, characterised by the effective lifetime~\cite{Fleischer:2011cw}.
For decay to a \CP\ eigenstate, in the absence of \CP\ violation in decay, the effective lifetime is the inverse of $\Gamma \pm \costbs \Delta\Gamma/2$ where the sign is the same as the \CP\ eigenvalue.  
It is thus interesting to ask if weighting the untagged data changes the decay-time distribution.  
To investigate this, it is useful to expand
\begin{equation}
\label{eq:Re-part}
    -2\mathcal{R}e\Big{(}e^{i2\beta_s}\mathcal{A}_f^*\overline{\mathcal{A}}_f\Big{)} = 
    -2 \left( \mathcal{R}e\left(\mathcal{A}_f^*\overline{\mathcal{A}}_f\right)\costbs - \mathcal{I}m\left(\mathcal{A}_f^*\overline{\mathcal{A}}_f\right)\sintbs \right) \,.
\end{equation}
Assuming absence of \CP\ violation in decay, the term in Eq.~\eqref{eq:Re-part} that is symmetric (antisymmetric) across the phase-space is sensitive to \costbs\ (\sintbs).
This is in contrast to Eq.~\eqref{eq:Im-part}, where it is the symmetric (antisymmetric) term that is sensitive to the sine (cosine) of the relevant weak phase $\phi_{\rm mix+dec}$.

The unweighted, untagged decay-time distribution is the sum of the \Bs\ and \Bsb\ decay rates, integrated across the phase space,
\begin{eqnarray}\label{eq:Bs-untagged}
    \Gamma_{\rm untagged}(t) & \propto & \\
    && \hspace{-20mm} e^{-\Gamma t}\left( 
        \int_{\rm DP}\left[ |\mathcal{A}_f|^2+|\overline{\mathcal{A}}_f|^2 \right] \cosh\left(\frac{\Delta\Gamma t}{2}\right) -2\mathcal{R}e\Big{(}e^{i2\beta_s}\mathcal{A}_f^*\overline{\mathcal{A}}_f\Big{)} \sinh\left(\frac{\Delta\Gamma t}{2}\right)\,{\rm d}\Omega
    \right)\,. \nonumber   
\end{eqnarray}
A Taylor expansion in $t$ can be used to identify the coefficient of the first-order term as the inverse of the effective lifetime, 
\begin{eqnarray}
    \Gamma_{\rm untagged}(t) & \appropto & 
    \left( 1 - \Gamma t \right) \left( \int_{\rm DP}|\mathcal{A}_f|^2+|\overline{\mathcal{A}}_f|^2 - 2\mathcal{R}e\Big{(}e^{i2\beta_s}\mathcal{A}_f^*\overline{\mathcal{A}}_f\Big{)} \left(\frac{\Delta\Gamma t}{2}\right) {\rm d}\Omega \right) \,, \nonumber \\
    & \appropto & \left( 1 - \Gamma t \right) \left( 1 - \frac{\int_{\rm DP}2\mathcal{R}e\Big{(}e^{i2\beta_s}\mathcal{A}_f^*\overline{\mathcal{A}}_f\Big{)}{\rm d}\Omega}{\int_{\rm DP}|\mathcal{A}_f|^2+|\overline{\mathcal{A}}_f|^2\,{\rm d}\Omega} \frac{\Delta\Gamma t}{2} \right) \,, \nonumber \\
    & \appropto & 1 - \left\{ \Gamma + \frac{\int_{\rm DP}2\mathcal{R}e\Big{(}e^{i2\beta_s}\mathcal{A}_f^*\overline{\mathcal{A}}_f\Big{)}{\rm d}\Omega}{\int_{\rm DP}|\mathcal{A}_f|^2+|\overline{\mathcal{A}}_f|^2\,{\rm d}\Omega} \frac{\Delta\Gamma}{2}\right\} t \,, \nonumber \\
    & \appropto & 1 - \left\{ \Gamma + \costbs \frac{\int_{\rm DP}2 \mathcal{R}e\Big{(}\mathcal{A}_f^*\overline{\mathcal{A}}_f\Big{)}{\rm d}\Omega}{\int_{\rm DP}\left( |\mathcal{A}_f|^2+|\overline{\mathcal{A}}_f|^2 \right)\,{\rm d}\Omega} \frac{\Delta\Gamma}{2}\right\} t \,, \label{eq:Gamma-untagged-unweighted}
\end{eqnarray}
where in the last line only the term that is symmetric across the phase space, and hence does not vanish on integration, is retained.
Here, and in similar Taylor expansions below, terms of ${\cal O}(t^2)$ and higher within each set of parentheses are omitted.
In the case of a pure \CP\ eigenstate, Eq.~\eqref{eq:Gamma-untagged-unweighted} gives the same effective lifetime as discussed above.
For a final state that is not a pure \CP\ eigenstate, the deviation of the effective lifetime from $\Gamma$ will be diluted by the hadronic factor in the last line of Eq.~\eqref{eq:Gamma-untagged-unweighted}, which is bound in the range $\left[-1,+1\right]$.\footnote{
    This hadronic factor, $\int_{\rm DP} 2\,\mathcal{R}e(\mathcal{A}_f^*\overline{\mathcal{A}}_f) {\rm d}\Omega/\int_{\rm DP} \left(|\mathcal{A}_f|^2+|\overline{\mathcal{A}}_f|^2\right) {\rm d}\Omega$, corresponds to the ``net \CP'' of a multibody final state, 
    which has been noted previously to be of particular relevance for neutral charm meson decays when used in $B\to DK$ processes to determine the CKM angle $\gamma$~\cite{BaBar:2007dro,Nayak:2014tea,Gershon:2015xra}.
}

Consider now the untagged decay-time distribution obtained after weighting with the $w^{\mathcal{R}e}$ function of Eq.~\eqref{eq:weighting-functions}, integrating across the phase-space, and Taylor-expanding in $t$,
\small
\begin{eqnarray}
    \Gamma^{w-\mathcal{R}e}_{\rm untagged}(t) & \propto & \nonumber \\
    && \hspace{-20mm} e^{-\Gamma t}\left( 
        \int_{\rm DP}2\,\mathcal{R}e(\mathcal{A}_f^*\overline{\mathcal{A}}_f) \cosh\left(\frac{\Delta\Gamma t}{2}\right) 
        -4 \costbs \frac{\left[\mathcal{R}e\left(\mathcal{A}_f^*\overline{\mathcal{A}}_f\right)\right]^2}{\left(|\mathcal{A}_f|^2+|\overline{\mathcal{A}}_f|^2\right)}\sinh\left(\frac{\Delta\Gamma t}{2}\right)\,{\rm d}\Omega
    \right)\,, \nonumber  \\ 
    & \appropto & \left( 1 - \Gamma t \right) \left( 
        \int_{\rm DP}2\,\mathcal{R}e(\mathcal{A}_f^*\overline{\mathcal{A}}_f) 
        -4 \costbs \frac{\left[\mathcal{R}e\left(\mathcal{A}_f^*\overline{\mathcal{A}}_f\right)\right]^2}{\left(|\mathcal{A}_f|^2+|\overline{\mathcal{A}}_f|^2\right)}\left(\frac{\Delta\Gamma t}{2}\right)\,{\rm d}\Omega
    \right)\,, \nonumber  \\ 
    & \appropto & 1 - \left\{ \Gamma + \costbs \frac{\int_{\rm DP}2\left[\mathcal{R}e\left(\mathcal{A}_f^*\overline{\mathcal{A}}_f\right)\right]^2/\left(|\mathcal{A}_f|^2+|\overline{\mathcal{A}}_f|^2\right)\,{\rm d}\Omega}{\int_{\rm DP}\mathcal{R}e(\mathcal{A}_f^*\overline{\mathcal{A}}_f)\,{\rm d}\Omega} \frac{\Delta\Gamma}{2} \right\} t \,. \label{eq:Gamma-untagged-wRe}
\end{eqnarray}
\normalsize
In the case of a pure \CP\ eigenstate this again gives the same result,\footnote{
    For a \CP-odd state, $w^{\mathcal{R}e}$ is equal to $-1$, so the weighted untagged decay-time distribution is flipped negative but otherwise unchanged.
    This sign flip is cancelled out by the division in the last line of Eq.~\eqref{eq:Gamma-untagged-wRe}.
} but in general a modification of the unweighted effective lifetime can be anticipated. 
This modification can be significant, since the $w^{\mathcal{R}e}$ function can be negative as well as positive.
Consequently, the coefficient of the $\cosh\left(\frac{\Delta\Gamma t}{2}\right)$ term in the first line of Eq.~\eqref{eq:Gamma-untagged-wRe} can also be either positive or negative, and can also be equal to zero in the limit that the final state is an equal mixture of \CP-even and \CP-odd (\ie\ has net-\CP of zero).
In this limit, the division in the last line of Eq.~\eqref{eq:Gamma-untagged-wRe} cannot be made and the distribution cannot be approximated by an exponential function.
Nonetheless, the dependence of the $w^{\mathcal{R}e}$-weighted decay-time distribution of Eq.~\eqref{eq:Gamma-untagged-wRe} on the weak phase factor is only through \costbs, as in Eq.~\eqref{eq:Gamma-untagged-unweighted}.

Turning now to the untagged decay-time distribution obtained after weighting with the $w^{\mathcal{I}m}$ function of Eq.~\eqref{eq:weighting-functions}, integrating across the phase-space, and Taylor-expanding in $t$,
\begin{eqnarray}
    \Gamma^{w-\mathcal{I}m}_{\rm untagged}(t) & \propto & e^{-\Gamma t}\left( 
        \int_{\rm DP}4 \sintbs \frac{\left[\mathcal{I}m\left(\mathcal{A}_f^*\overline{\mathcal{A}}_f\right)\right]^2}{\left(|\mathcal{A}_f|^2+|\overline{\mathcal{A}}_f|^2\right)}\sinh\left(\frac{\Delta\Gamma t}{2}\right)\,{\rm d}\Omega
    \right)\,, \nonumber  \\ 
    & \appropto & \left( 1 - \Gamma t \right) \left( 
        \int_{\rm DP}4 \sintbs \frac{\left[\mathcal{I}m\left(\mathcal{A}_f^*\overline{\mathcal{A}}_f\right)\right]^2}{\left(|\mathcal{A}_f|^2+|\overline{\mathcal{A}}_f|^2\right)}\left(\frac{\Delta\Gamma t}{2}\right)\,{\rm d}\Omega
    \right)\,, \nonumber  \\ 
    & \appropto & \left\{ \sintbs \frac{\int_{\rm DP}4\left[\mathcal{I}m\left(\mathcal{A}_f^*\overline{\mathcal{A}}_f\right)\right]^2/\left(|\mathcal{A}_f|^2+|\overline{\mathcal{A}}_f|^2\right)\,{\rm d}\Omega}{\int_{\rm DP}\left( |\mathcal{A}_f|^2+|\overline{\mathcal{A}}_f|^2 \right)\,{\rm d}\Omega} \frac{\Delta \Gamma}{2} \right\} t\,. \label{eq:Gamma-untagged-wIm}
\end{eqnarray}
In the last line, the hadronic factor has been written in a way that makes it independent of the normalisation of $\mathcal{A}_f$ and $\overline{\mathcal{A}}_f$, as done previously in Eqs.~\eqref{eq:Gamma-untagged-unweighted} and~\eqref{eq:Gamma-untagged-wRe}.
Due to the cancellation of the $\cosh(\Delta\Gamma/2)$ term, the lowest-order term in the Taylor expansion in Eq.~\eqref{eq:Gamma-untagged-wIm} is linear in $t$, rather than constant, and thus this expression cannot be approximated by a single exponential function.\footnote{
    Clearly the lowest order Taylor expansion is only a good approximation at values of $t$ satisfying $\Gamma t \ll 1$.
    It is nonetheless useful to illustrate the differences between the unweighted and weighted untagged decay-time distributions.
}
The magnitude of $\Gamma^{w-\mathcal{I}m}_{\rm untagged}(t)$ is proportional to \sintbs, and thus corresponds to \CP\ violation in the interference between mixing and decay in the \Bs\ system.  
Although the $w^{\mathcal{I}m}$ function can be both positive and negative, all hadronic factors in Eq.~\eqref{eq:Gamma-untagged-wIm} are positive, and since $\Delta\Gamma$ is also known to be positive the sign of $\Gamma^{w-\mathcal{I}m}_{\rm untagged}(t)$ depends only on the sign of \sintbs.

Effectively, the weighted untagged decay-time distribution of Eq.~\eqref{eq:Gamma-untagged-wIm} allows visualisation of the emergence, as a function of decay time, of an asymmetry across the Dalitz-plot.
It therefore shares some similarity with methods of observing \CP\ violation in decay through (decay-time-integrated) Dalitz-plot asymmetries~\cite{Gardner:2002bb,Gardner:2003su}, but is novel in that it allows \CP\ violation in the interference between mixing and decay to be observed.
As this method requires neither flavour tagging nor ability to resolve the fast $\Bs$--$\Bsb$ oscillations, it may enable some experiments to achieve sensitivity to \sintbs\ that would otherwise be impossible.\footnote{
    The potential to determine \sintbs\ from an untagged analysis of $\Bs \to \jpsi\phi$ decays was noted in Ref.~\cite{Dunietz:2000cr}, and the method was implemented by the D0 collaboration in Refs.~\cite{D0:2007ewb,D0:2007jut}, but the use of weighting to visualise the effect has not been described previously in the literature.
}

Normalisation has been omitted in Eqs.~\eqref{eq:Bs-untagged}--\eqref{eq:Gamma-untagged-wIm} for simplicity, even though the argument made in Sec.~\ref{sec:methodology} that it is irrelevant due to cancellation in the asymmetries no longer holds.  
For a useful visualisation of data, with fit curve superimposed, the key issue is that the same normalisation should be used for both.
In practice it is likely to be convenient to normalise Eq.~\eqref{eq:Bs-untagged} to unity, \ie\ to treat it as a probability density function, and to use the same normalisation factor also in Eqs.~\eqref{eq:Gamma-untagged-wRe} and~\eqref{eq:Gamma-untagged-wIm}.
Since both of these weighted untagged decay-time distributions could, in principle, be equal to zero, they cannot in general be normalised individually.

To illustrate the concept of this weighted untagged decay-time distribution, the model for $\Bz \to D_{\CP}\pip\pim$ decays described in Sec.~\ref{sec:illustration} is modified to introduce a decay-width difference.  
The value of the decay-width difference is taken to be that of the $\Bs$--$\Bsb$ system, specifically $\Delta \Gamma = 0.083 \ps^{-1}$~\cite{HFLAV21}, but all other features of the model remain the same (\eg\ the mass, lifetime, mass difference and weak phase factor are those of the $\Bz$--$\Bzb$ system).
This has the advantage that the larger magnitude of $\sintb$ compared to $\sintbs$ provides a more easily visible effect in Eq.~\eqref{eq:Gamma-untagged-wIm}.
The hadronic factors of Eqs.~\eqref{eq:Gamma-untagged-unweighted}--\eqref{eq:Gamma-untagged-wIm} are evaluated with this decay model to be
\begin{eqnarray}
    \label{eq:wfactor4}
    \frac{\int_{\rm DP}2 \mathcal{R}e\Big{(}\mathcal{A}_f^*\overline{\mathcal{A}}_f\Big{)}{\rm d}\Omega}{\int_{\rm DP}\left( |\mathcal{A}_f|^2+|\overline{\mathcal{A}}_f|^2 \right)\,{\rm d}\Omega} & = & -0.085\,, \\
    \label{eq:wfactor5}
    \frac{\int_{\rm DP}2\left[\mathcal{R}e\left(\mathcal{A}_f^*\overline{\mathcal{A}}_f\right)\right]^2/\left(|\mathcal{A}_f|^2+|\overline{\mathcal{A}}_f|^2\right)\,{\rm d}\Omega}{\int_{\rm DP}\mathcal{R}e(\mathcal{A}_f^*\overline{\mathcal{A}}_f)\,{\rm d}\Omega} & = & -3.121\,, \\
    \label{eq:wfactor6}
    \frac{\int_{\rm DP}4\left[\mathcal{I}m\left(\mathcal{A}_f^*\overline{\mathcal{A}}_f\right)\right]^2/\left(|\mathcal{A}_f|^2+|\overline{\mathcal{A}}_f|^2\right)\,{\rm d}\Omega}{\int_{\rm DP}\left( |\mathcal{A}_f|^2+|\overline{\mathcal{A}}_f|^2 \right)\,{\rm d}\Omega} & = & \phantom{-}0.229 \, ,
\end{eqnarray}
where the last is the same as given in Eq.~\eqref{eq:wfactor2}.

\begin{figure}[tb]
    \centering
    \includegraphics[width=0.48\linewidth]{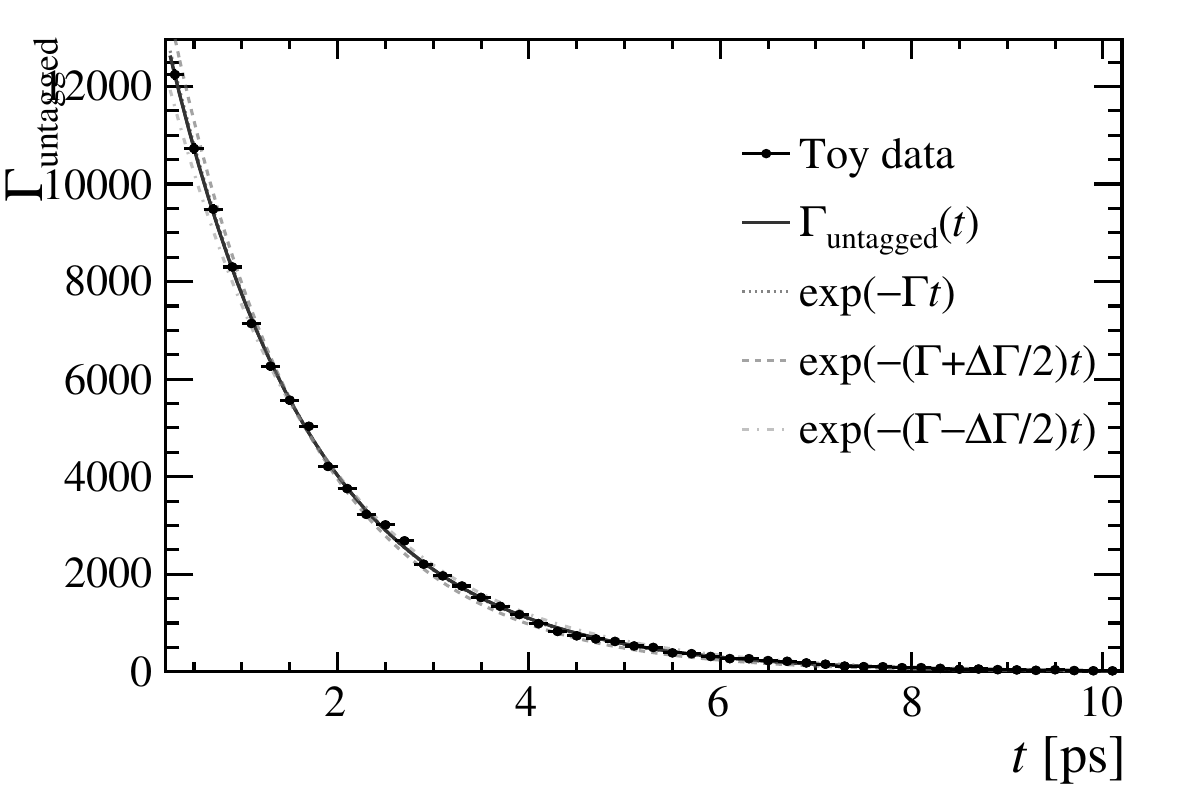} \\
    \includegraphics[width=0.48\linewidth]{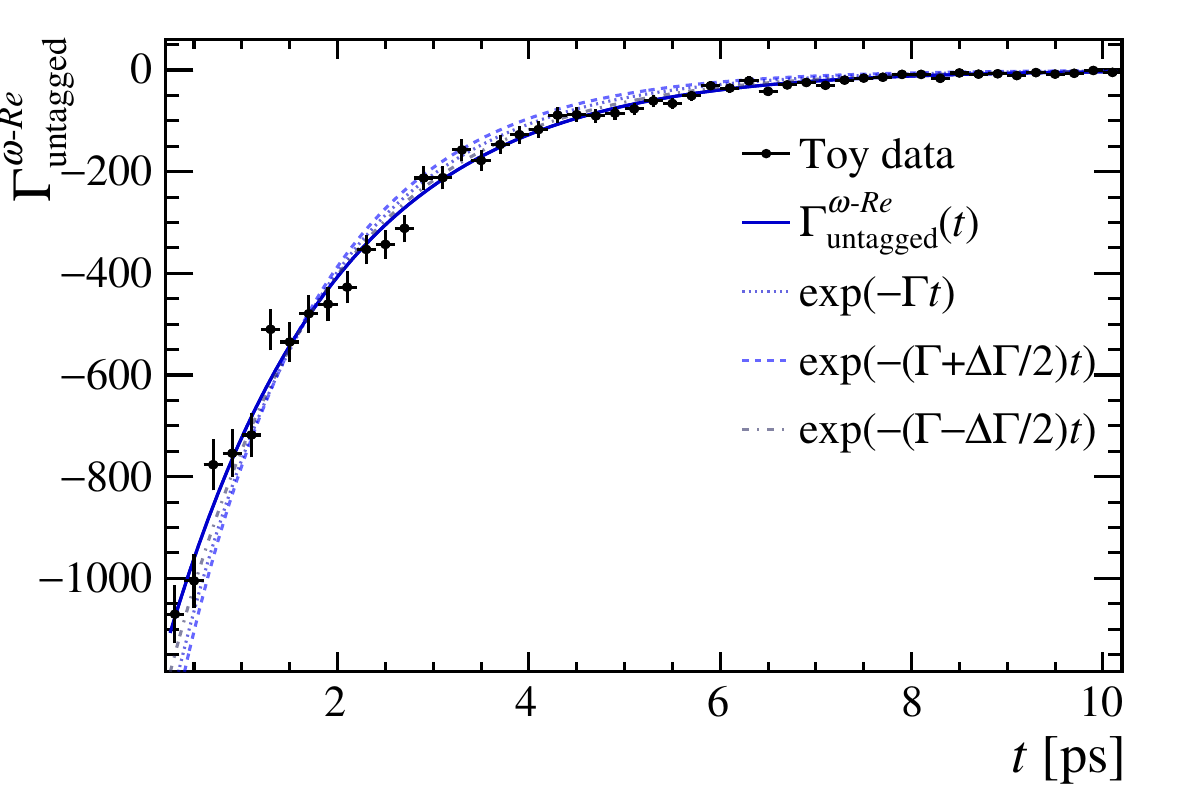}
    \includegraphics[width=0.48\linewidth]{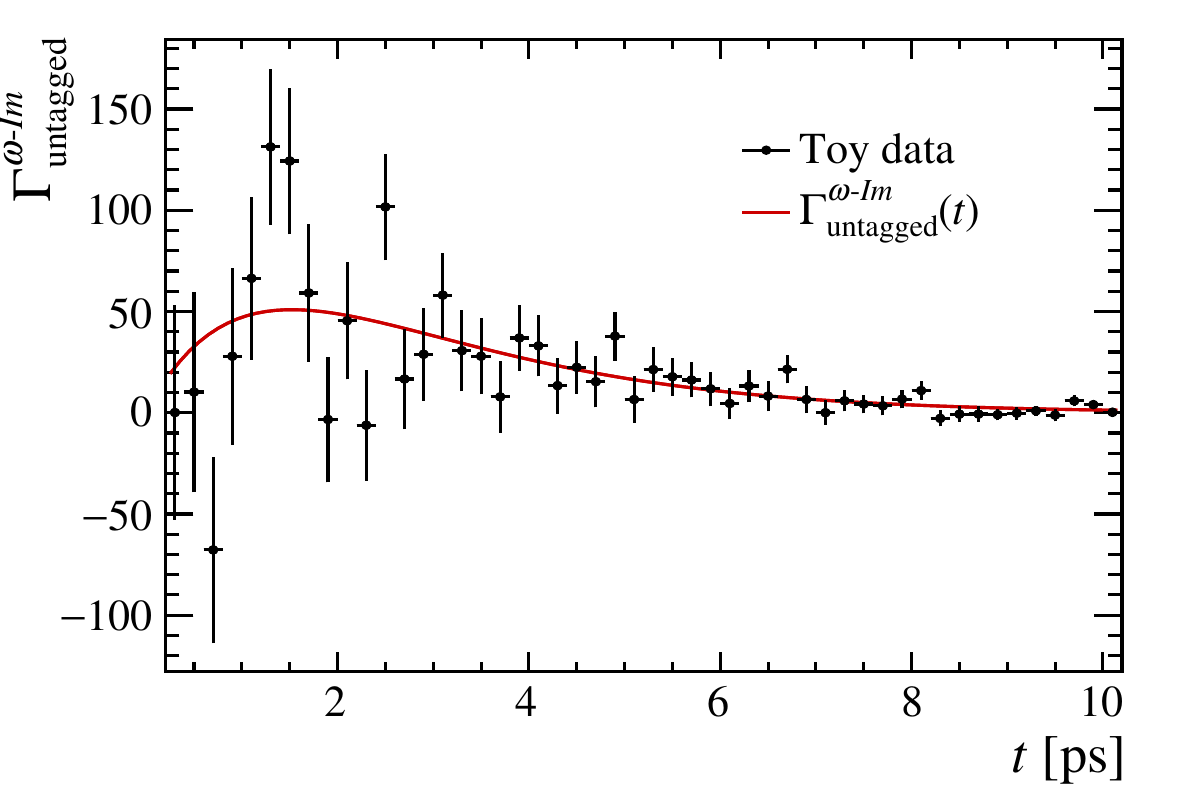}
    \caption{Distributions of untagged decay-rate distributions (top)~without weighting, and with (bottom left)~$w^{\mathcal{R}e}$ and (bottom right)~$w^{\mathcal{I}m}$ weights applied.
    These are obtained from pseudoexperiments generated with the Dalitz-plot model from Ref.~\cite{LHCb-PAPER-2014-070}, with a non-zero width difference introduced as described in the text.
    Curves corresponding to the expected distributions without approximations, from Eqs.~\eqref{eq:Bs-untagged},~\eqref{eq:Gamma-untagged-wRe} and~\eqref{eq:Gamma-untagged-wIm}, are also shown and compared to exponential distributions with different decay widths.}
    \label{fig:wGammaUntagged-toy}
\end{figure}

The untagged decay-rate distributions obtained from a sample generated with this model, both without weighting and with the $w^{\mathcal{R}e}$ and $w^{\mathcal{I}m}$ weighting schemes are shown in Fig.~\ref{fig:wGammaUntagged-toy}.
A large sample of $100\,000$ events is used to illustrate the impact of the method.
Since the net \CP\ of the $\Bz \to D_{\CP}\pip\pim$ model, Eq.~\eqref{eq:wfactor4}, is close to zero, the unweighted untagged decay-time distribution is well approximated by an exponential decay with the \Bz\ lifetime.
The hadronic factor of Eq.~\eqref{eq:wfactor5} is, on the other hand, significantly different from zero, causing the $w^{\mathcal{R}e}$-weighted distribution to have a different effective lifetime (as well as being flipped negative).
The modest value of the hadronic factor of Eq.~\eqref{eq:wfactor6} allows the term corresponding to \CP-violation in the interference of mixing and decay to be seen.

There are a range of \Bs\ decays where this visualisation method may be useful in future.  
The most extensively studied to date is $\Bs\to \jpsi\Kp\Km$~\cite{LHCb-PAPER-2023-016,LHCb-PAPER-2017-008,ATLAS:2020lbz,CMS:2020efq}, for which it will be important to be able to see in the data the oscillations caused by a non-zero value of \sintbs\ as the sensitivity approaches the level at which this is likely to be observed.
As noted above, the $w^{\mathcal{I}m}$-weighted untagged decay-time distribution may be particularly useful for experiments where the decay-time resolution is not sufficient to resolve the fast $\Bs$--$\Bsb$ oscillations.
There could also be interesting applications of the method in the study of $\Bs \to D\Kp\Km$ decays, where the $D$ meson can be reconstructed in a number of different final states, which is of interest to determine the CKM angle $\gamma$~\cite{Dunietz:1995cp,Gronau:2004gt,Gronau:2007bh,Ao:2020cwh}.
Another potentially interesting mode is $\Bs \to \KS\pip\pim$, which provides an alternative way to determine $\gamma$~\cite{Ciuchini:2006st,Gershon:2014yma}.
Consideration of the impact of \CP\ violation in decay may be important in these cases, however.

\section{\texorpdfstring{\boldmath Impact of \CP\ violation in decay}{Impact of CP violation in decay}}
\label{sec:CPVinDecay}

The visualisation method discussed so far is applicable for \Bz-meson decays to multibody, self-conjugate final states where there is no \CP\ violation in decay.  
This is the case, to a good approximation, for a number of decay modes of interest, including $\Bz \to \jpsi \Kstar(892)$ with $\Kstar(892) \to \KS\piz$, $\Bd \to D^{(*)}h^0$ with $D \to \KS\pip\pim$, $\Bz \to \jpsi \pip\pim$ and $\Bz \to D_{\CP}\pip\pim$.
In these cases it may nonetheless become important, as sample sizes increase, to be able to assess the impact of non-zero \CP\ violation effects.
Moreover, it would be of interest to apply the method also to decays where large \CP\ violation effects are either known or expected, including $\Bz \to \Kp\Km\KS$, $\Bz \to \KS\pip\pim$ and $\Bz \to \pip\pim\piz$.

The presence of \CP\ violation in decay implies that the amplitude for decay to a final state that is, say, \CP-even must include a contribution from the \CP-odd $B$ state.
This breaks the interpretation of Eqs.~\eqref{eq:symRel-Re} and~\eqref{eq:symRel-Im} in terms of symmetries across the phase space --- it means that $A_{+\,f}$ ($A_{-\,f}$) is no longer symmetric (antisymmetric) about the Dalitz-plot symmetry line $m_+^2 = m_-^2$.
Thus, application of the $w^{\mathcal{R}e}$ and $w^{\mathcal{I}m}$ weighting functions defined in Eq.~\eqref{eq:weighting-functions} will no longer lead to the cancellation of terms that simplifies the asymmetries given in Eqs.~\eqref{eq:wacp_re_def}--\eqref{eq:wC_im_def}.
Additionally, the presence of \CP\ violation in decay means that weak phase factors cannot in general be separated from hadronic factors.
While in Eq.~\eqref{eq:ampt} the weak phase ($-2\beta$) could be separated from the amplitudes $\mathcal{A}_f$ and $\overline{\mathcal{A}}_f$, the existence of \CP\ violation in decay means that $\mathcal{A}_f$ and $\overline{\mathcal{A}}_f$ must be composed from two contributions with different weak phases (as well as a non-trivial strong phase difference).
Since the relative magnitudes of these two contributions would be expected to vary across the phase-space, due to the presence of resonances, the weak phase difference between $\mathcal{A}_f$ and $\overline{\mathcal{A}}_f$ will also vary across the phase-space and therefore cannot in general be factored out as a constant term.

The above observations are related to the well-known facts that for decays to \CP\ eigenstates the presence of \CP\ violation in decay introduces a cosine term to the decay-time-dependent asymmetry, and precludes a straightforward relation of the coefficient of the sine term to a weak phase. 
Nonetheless, as is the case for decays to \CP\ eigenstates, the weighted asymmetries may provide an interesting and useful visualisation method, as it will still be possible to compare the effects observed in data to those expected in a model, including when the model is the result of a fit to data.

\section{Summary}
\label{sec:summary}

A new method to weight data in order to visualise decay-time-dependent \CP\ violation effects in neutral \B\ meson decays to multibody final states, that are not \CP\ eigenstates, is proposed.
In the common case that the weak phase difference between \Bz\ and \Bzb\ decay amplitudes, also accounting for the phase from $\Bz$--$\Bzb$ mixing, is $-2\beta$ and that effects of \CP\ violation in decay can be neglected, the method allows separate visualisation of asymmetries proportional to \sintb\ and \costb.
The method is based on fundamental symmetries of the decay amplitudes, and is applicable to a wide range of decay channels.
Weighted untagged decay-time distributions can also be formed, and provide novel potential for observation of effects due to \CP\ violation in the interference between mixing and decay in the \Bs\ system, exploiting the non-zero decay-width difference.

%% file: ack.tex
\section*{Acknowledgements}

The authors wish to thank their colleagues on the LHCb experiment for the fruitful and enjoyable collaboration that inspired this study.
In particular, they would like to thank Matthew Kenzie and Peilian Li for helpful comments.
Additionally, the authors acknowledge stimulating discussions with Yasmine Amhis, Peter Clarke,  Juli{\'a}n Garc{\'\i}a Pardi{\~n}as and Stephane Monteil.
TG, TL and MW are supported by the Science and Technology Facilities Council (UK).
AM acknowledges support from the European Research Council Starting grant ALPACA 101040710. 
WQ is supported by the National Science Foundation of China under Grant No.\ 11975015, the National Key R\&D Program of China under Grant No.\ 2022YFA1601901 and Fundamental Research Funds for the Central Universities.